\documentclass[reqno,11pt]{amsart}
\usepackage{amsthm}
\usepackage{amssymb}
\usepackage{amsmath}
\usepackage{hyperref}
\usepackage[mathscr]{eucal}
\usepackage{enumerate}

\numberwithin{equation}{section}

\newtheorem{Def}{Definition}[section]
\newtheorem{Thm}[Def]{Theorem}
\newtheorem{Prop}[Def]{Proposition}
\newtheorem{Lemma}[Def]{Lemma}

\newcommand{\beq}{\begin{equation}}
\newcommand{\eeq}{\end{equation}}

\title[The Riemann-flat Condition]{Shock Wave Interactions and the Riemann-flat Condition: The Geometry behind Metric Smoothing and the Existence of Locally Inertial Frames in General Relativity}

\author[M.\ Reintjes]{Moritz Reintjes}
\address{Fachbereich f\"ur Mathematik \\ Universit\"at Konstanz \\ 78464 Konstanz \\ Germany}
\email{moritzreintjes@gmail.com}

\author[B.\ Temple]{Blake Temple \\ \\ August 18, 2019}
\address{Department of Mathematics\\ University of California\\ Davis, CA 95616\\ USA}
 \email{temple@math.ucdavis.edu}

\begin{document}

\maketitle

\begin{abstract} 
We prove that the essential smoothness of the gravitational metric at shock waves in GR, a PDE regularity issue for weak solutions of the Einstein equations, is determined by a geometrical condition which we introduce and name the {\it Riemann-flat condition}. The Riemann-flat condition determines whether or not the essential smoothness of the gravitational metric is two full derivatives more regular than the Riemann curvature tensor.  This provides a geometric framework for the open problem as to whether {\it regularity singularities} (points where the curvature is in $L^\infty$ but the essential smoothness of the gravitational metric is only Lipschitz continuous) can be created by shock wave interaction in GR, or whether metrics Lipschitz at shocks can always be smoothed one level to $C^{1,1}$ by coordinate transformation.   As a corollary of the ideas we give a proof that locally inertial frames always exist in a natural sense for shock wave metrics in spherically symmetric spacetimes, independent of whether the metric itself can be smoothed to $C^{1,1}$ locally.  This latter result yields an explicit procedure (analogous to Riemann Normal Coordinates in smooth spacetimes) for constructing locally inertial coordinates for Lipschitz metrics, and is a new regularity result for GR solutions constructed by the Glimm scheme.  
\end{abstract}

\section{Introduction}\label{intro}

We introduce the {\it Riemann-flat condition} on a spacetime connection $\Gamma$ and prove that this condition is necessary and sufficient for determining the essential smoothness of weak solutions of the Einstein equations at apparent singularities where the gravitational metric tensor $g$ is only Lipschitz continuous $C^{0,1}$.  The condition applies in the general setting of connections $\Gamma\in L^{\infty}$, under the assumption that the curvature tensor ${\rm Riem}(\Gamma)$ has the same regularity as the connection, ${\rm Riem}(\Gamma)\in L^{\infty}$, a natural framework for shock wave solutions in GR.   The Riemann-flat condition is the condition that there exist a Lipschitz tensor $\tilde{\Gamma}$ such that the associated connection $\Gamma-\tilde{\Gamma}$ is Riemann-flat, and we prove that there exists a coordinate transformation within the $C^{1,1}$ atlas which smooths the metric components from $C^{0,1}$ to $C^{1,1}$ if and only if $\Gamma$ is Riemann-flat, (c.f. Theorem \ref{thm_equivalence} below).\footnote{The space $C^{0,1}$ denotes the space of Lipschitz continuous functions, and $C^{1,1}$ the space of functions with Lipschitz continuous derivatives. A function is bounded in $C^{0,1}$ if and only if the function and its weak derivatives are bounded in $L^\infty$, c.f. \cite{Evans}, Chapter 5.8.} Theorem \ref{thm_equivalence} applies at points of arbitrarily complex shock wave interactions, in $n$-dimensions, without assuming any spacetime symmetries or technical assumptions. The space of $L^{\infty}$ connections with $L^{\infty}$ Riemann curvature tensor is closed under $C^{1,1}$ coordinate transformations, so the $C^{1,1}$ atlas is natural for shock wave theory in GR.\footnote{In geometry, the regularity of a Riemannian or Lorentzian metric is defined to be the regularity of the metric components in each coordinate system of a given atlas. This definition neglects the possibility that the metric might be more regular in particular coordinate systems of this atlas.}  

To prove Theorem \ref{thm_equivalence}, that a connection $\Gamma \in L^\infty$ which meets the Riemann-flat condition can always be smoothed to $C^{0,1}$ by a $C^{1,1}$ coordinate transformation, the main step is to prove that if the Riemann curvature tensor is {\it zero} in the weak sense, then there exists a $C^{1,1}$ coordinate transformation that maps the connection to zero, even if the connection is only in $L^{\infty}$.   That is, the main step in proving the equivalence of the Riemann-flat condition for connections of nonzero curvature, is to resolve the problem of metric smoothing in the special case when the curvature of the connection is zero.  As a corollary of the ideas we exhibit an explicit construction procedure and proof that locally inertial frames exist in a natural sense at points of arbitrary shock wave interaction in spherically symmetric spacetimes when the gravitational metric is only Lipschitz continuous.   This establishes that the $C^{0,1}$ shock wave metrics generated by the Glimm scheme in \cite{GroahTemple}, are locally inertial at every point,  independent of whether the metric can be smoothed locally to $C^{1,1}$.   This new result is stated below in Theorem \ref{thm_noRS} of Section \ref{sec_results}.

One could interpret Theorem \ref{thm_equivalence} in the spirit of the Nash embedding theorems \cite{Nash}. Namely, since the addition of a Lipschitz tensor would not alter the discontinuities across shocks which form the singular set of $\Gamma$, Theorem \ref{thm_equivalence} states that one can smooth the connection if and only if there exists a Riemann-flat $L^{\infty}$ connection $\hat{\Gamma}=\Gamma-\tilde{\Gamma}$ which has the same jump discontinuities as the original connection $\Gamma$ on the same singular set, because $\tilde{\Gamma}$ is a continuous function.  Thus since $\hat{\Gamma}$ is flat, it can be interpreted as an extension of the singular part of $\Gamma$ into flat space, so the open question of regularity singularities can be thought of as whether one can embed the singular part of $\Gamma$ into ambient flat space without changing the jumps.\footnote{Since the writing of this paper the Riemann-flat condition has become the starting point for authors' further developments. In fact, we abandoned the Nash embedding idea and used the Riemann-flat condition to derive the \emph{Regularity Transformation equations}, an elliptic system of PDE's equivalent to the Riemann-flat condition, c.f. \cite{ReintjesTemple3,ReintjesTemple4,ReintjesTemple5}.}

\section{Motivation and Background}\label{motbac}

It is well known that shock waves form in solutions of the Einstein-Euler equations, the equations that couple the spacetime geometry to perfect fluid sources, whenever the flow is sufficiently compressive \cite{Christodoulou,Lax,Smoller}. But it is an open question as to the essential level of smoothness of the gravitational metric for general shock wave solutions admitting points of shock wave interaction.                       The existence theory \cite{GroahTemple} for shock waves in GR based on the Glimm scheme, (see also \cite{LeFlochStewart}), only yields Lipschitz continuity of the spacetime metric, a regularity too low to guarantee the existence of locally inertial coordinates within the atlas of {\it smooth} ($C^2$) coordinate transformations \cite{ReintjesTemple1}.  That spacetime is locally inertial at each point $p$, (i.e., there exists coordinate systems in which the metric is Minkowski at $p$, and all coordinate derivatives of the metric vanish at $p$), was Einstein's starting assumption for General Relativity, \cite{Einstein}.   The requisite smoothness of the metric sufficient to guarantee the existence of locally inertial frames within the {\it smooth} atlas, is the metric regularity $C^{1,1}$, one degree smoother than the $C^{0,1}$ metrics constructed in \cite{GroahTemple}. In the smooth case, the Riemann normal coordinate construction generates a smooth transformation to locally inertial coordinates. In \cite{ReintjesTemple1}, the authors proposed the possibility that shock wave interaction might create a new kind of spacetime singularity which we named {\it regularity singularities},  a point in spacetime where the gravitational metric tensor is Lipschitz continuous, but essentially less regular than $C^{1,1}$.

Like other singularities in GR, such as the event horizon of the Schwarz\-schild spacetime, a singularity requires a \emph{singular} coordinate transformation to regularize it. Thus the possibility remains that the spacetime metric at shock waves might be smoothed from $C^{0,1}$ to $C^{1,1}$ within the larger atlas of less regular $C^{1,1}$ coordinate transformations, because these transformations introduce jumps in the derivatives of the Jacobian which hold the potential to eliminate the jumps in metric derivatives.  It remains an outstanding open problem as to whether such transformations exist to smooth the metric to $C^{1,1}$ at points of shock wave interaction in GR.   If such smoothing transformations do not exist, then  regularity singularities can be created by shock wave interaction alone. In particular, this would imply new scattering effects in gravitational radiation,  \cite{ReintjesTemple2}.

The starting point for addressing this basic regularity question for GR shock waves is Israel's celebrated 1966 paper \cite{Israel}, which proves that, for any smooth co-dimension one shock surface in $n$ dimensions, the gravitational metric can always be smoothed from $C^{0,1}$ to $C^{1,1}$ by transformation to Gaussian normal coordinates adjusted to the shock surface. This transformation was identified as an element of the $C^{1,1}$ atlas in \cite{SmollerTemple}. However, these coordinates are only defined for single, non-interacting shock surfaces and do not exist for the $C^{0,1}$ metrics containing shock wave \emph{interactions} constructed by the Glimm scheme in the Groah-Temple framework \cite{GroahTemple}.  The only result going beyond Israel's result to shock wave interactions was accomplished in \cite{Reintjes}, where the first author proved that the gravitational metric {\it can} always be smoothed from $C^{0,1}$ to $C^{1,1}$ at a point of regular shock wave interaction between shocks from different characteristic families, in spherically symmetric spacetimes.  The proof is based on a surprisingly complicated new constructive method which analyzes non-local PDE's tailored to the structure of the shock-wave interaction.   It is not clear whether or how this proof could be extended to significantly more complicated interactions. For more complicated shock wave interactions in spherically symmetric spacetimes, and general asymmetric shock interactions in $(3+1)$-dimensions, the question as to the locally flat nature of space-time, or whether regularity singularities can be created by shock wave interactions, remains an open problem.

The atlas of $C^{1,1}$ coordinate transformations was introduced in \cite{SmollerTemple} as the natural atlas for shock wave metrics with $C^{0,1}$ regularity in GR, because $C^{1,1}$ coordinate transformations preserve the Lipschitz continuity of the metric, and map bounded discontinuous curvature tensors to bounded discontinuous curvature tensors. For perfect fluids, shock waves are weak solutions of the Einstein-Euler equations, $G=\kappa T$ coupled with $Div\: T=0$, where $G$ is the Einstein curvature tensor, $T$ is the energy-momentum tensor for a perfect fluid, and $\kappa$ is the coupling constant, c.f. \cite{Choquet,Weinberg}.   At shock waves, $T$ is discontinuous and contains no delta function sources, the latter distinguishing shock waves from surface layers. The Einstein equations then imply that the curvature tensor $G$ must also be free of delta function sources at shock waves, \cite{Israel}.   Since $G$ contains second derivatives of the gravitational metric $g$, it follows that all delta function sources in the second derivatives of $g$ must cancel out to make $G$ bounded and discontinuous at the shocks. The results in \cite{SmollerTemple} prove that this cancellation of delta function sources is a covariant property within the $C^{1,1}$ atlas.   To rule out delta function sources in $G$, it is sufficient to assume the Riemann curvature tensor is bounded in $L^{\infty}$.\footnote{Note that there is no loss of generality in assuming the entire Riemann curvature tensor, not just the Einstein tensor $G$, is bounded in $L^{\infty}$, because the existence of delta function sources in the curvature tensor automatically prevents Lipschitz regularity of the connection.} 

The authors' work in \cite{ReintjesTemple2} indicated that the problem of the existence of locally inertial frames might be independent from the problem of the essential $C^{0,1}$ regularity of a general  connection.\footnote{For metric connections,  the Christoffel formulas give the connection in terms of first derivatives of the metric, so $C^{0,1}$ regularity of the connection is equivalent to $C^{1,1}$ regularity of the metric, \cite{Weinberg}.}   To make this distinction precise, we defined in \cite{ReintjesTemple2} a {\it regularity singularity} to be a point $p$ where the connection is essentially less smooth than $C^{0,1}$ in the sense that there does not exist a $C^{1,1}$ coordinate transformation in a neighborhood of $p$ that smooths the connection to $C^{0,1}$ in that neighborhood.  Independently, we say the $L^\infty$ connection is {\it locally inertial} at $p$,  if there exists a coordinate system within the $C^{1,1}$ atlas in which the connection vanishes at $p$, and is Lipschitz continuous just at $p$.  Thus, locally inertial coordinates could exist at $p$ even though the essential smoothness of the metric is less than $C^{0,1}$. Based on this, we say a regularity singularity at a point $p$ is {\it weak} if there exists a locally inertial coordinate system at $p$, and {\it strong} if the connection does not admit locally inertial coordinates at $p$. For example, at a weak regularity singularity in GR, locally inertial coordinates would exist at $p$, but the metric smoothness remains below $C^{1,1}$, too low for many desirable properties to hold, (e.g. the Penrose-Hawking-Ellis Singularity theorems \cite{HawkingEllis,GrafKunzinger}).\footnote{See \cite{Anderson,ChenLeFloch,Klainermann} for results on lower regularity solutions of the vacuum Einstein equations, a setting that rules out shock-waves.}  From this point of view,  Theorem 3 below establishes that GR solutions generated by the Glimm scheme cannot produce strong regularity singularities, but leaves open the possibility that weak regularity singularities still exist.

\section{Statement of Results} \label{sec_results}

Let $\mathcal{M}$ be an $n$-dimensional manifold endowed with a symmetric connection $\Gamma$ such that the components of $\Gamma$ and its curvature tensor $Riem(\Gamma)$ are bounded in $L^{\infty}$ in coordinate system $x$.  The space of $L^{\infty}$ connections with $L^{\infty}$ curvature tensors is invariant under $C^{1,1}$ coordinate transformations, \cite{SmollerTemple}, and provides a general covariant framework in which to address essential metric regularity at shock waves in GR.  Since we are interested in a local theory, assume $\Gamma$ is given in a fixed coordinate system $x^{i}$ defined in an open neighborhood $\mathcal{U}$ of a point $p$, and assume that in $x$-coordinates the connection components $\Gamma^{k}_{ij}$ satisfy
\begin{eqnarray}\label{sup}
\|\Gamma\|_\infty \equiv \max\limits_{k,i,j}\|\Gamma^{k}_{ij}\|_{L^\infty(\mathcal U)}\leq M_0,
\end{eqnarray}
for some constant $M_0>0$. We let $\mathcal{U}$ denote a neighborhood on the manifold and to reduce notation we use $\mathcal{U}$ to denote $\mathcal{U}$ as well. When $y$ coordinates are distinguished from $x$, we use the standard convention that components in $x$-coordinates use indices $i,j,k,...$ while components in $y$-coordinates use $\alpha,\beta,\gamma,...$. Our main theorem states that there exists a $C^{1,1}$ coordinate transformation that lifts the regularity of $\Gamma$ from $L^{\infty}$ to $C^{0,1}$ if and only if the connection meets the Riemann-flat condition, which we state first:

\begin{Def}\label{Riemflat} A symmetric connection $\Gamma\in L^{\infty}$ is said to meet the {\it Riemann-flat condition} at a point $p$ if there exists a symmetric Lipschitz continuous $(1,2)$-tensor $\tilde{\Gamma}^k_{ij}$ defined in a neighborhood of $p$, such that the connection  
\beq\label{hatGamma}
\hat{\Gamma}^k_{ij}\equiv \Gamma^k_{ij}- \tilde{\Gamma}^k_{ij}
\eeq 
satisfies 
${{\rm Riem}}(\hat{\Gamma})=0$ weakly in $L^\infty$, $($c.f. \eqref{curvature_weakform} below$)$. 
\end{Def}    

Note that a connection which satisfies the Riemann-flat condition automatically has Riemann curvature in $L^\infty$ because the derivative part of the curvature is linear.  Thus the assumption of bounded curvature is not explicitly needed in Definition \ref{Riemflat} and Theorem \ref{thm_equivalence} to follow.\footnote{It remains an open problem whether an $L^\infty$ connection always satisfies the Riemann-flat condition if its Riemann curvature tensor lies in $L^\infty$.}

\begin{Thm} \label{thm_equivalence} 
Assume $\Gamma^k_{ij}$ is a symmetric $L^{\infty}$ connection satisfying (\ref{sup}) in $x$-coordinates  defined in neighborhood $\mathcal{U}$ of a point $p\in\mathcal{M}$.  Then there exists a $C^{1,1}$ transformation $y\circ x^{-1}$ defined in a neighborhood of $p$, such that in $y$-coordinates
$
\Gamma^{\alpha}_{\beta\gamma} \in C^{0,1}
$
if and only if $\Gamma$ meets the Riemann-flat condition. 
Moreover, the smoothing transformation $y\circ x^{-1}$ is related to (\ref{hatGamma}) by 
\beq \label{formulafory}
\frac{\partial^2 y^\alpha}{\partial x^i \partial x^j} = \frac{\partial y^\alpha}{\partial x^k} \hat{\Gamma}^k_{ij};
\eeq 
and if either side of the equivalence holds we have 
\beq \label{Riem_bounded_Thm_equiv}
Riem(\Gamma) \in L^{\infty}(\mathcal{U}).
\eeq
\end{Thm}

The main step in proving Theorem \ref{thm_equivalence} is to establish the following proposition which asserts that connections with zero Riemann curvature can be mapped to zero (i.e., are Euclidean or Minkowski) by a coordinate transformation. This requires admitting a sufficiently singular atlas, which is $C^{1,1}$ when the connection is assumed in $L^{\infty}$.  
 
\begin{Prop}\label{thm1}  
Assume $\hat{\Gamma}^k_{ij}$ are the components of an $L^{\infty}$ symmetric connection $\hat{\Gamma}$ in $x$-coordinates defined in a neighborhood of $p\in\mathcal{M}$.  Then   
$
 {\rm Riem}(\hat{\Gamma})=0
$
in the $L^\infty$ weak sense in a neighborhood of $p$ if and only if there exists a $C^{1,1}$ transformation $y\circ x^{-1}$ such that, in $y$-coordinates,             
$
 \hat{\Gamma}^{\alpha}_{\beta\gamma}      \ = \ 0\ \ \ a.e.
$
in a neighborhood of $p$.   
\end{Prop}

Our second theorem uses ideas in the proof of Proposition \ref{thm1} to give a constructive proof that locally inertial frames always exist in a natural sense for the $C^{0,1}$ shock wave metrics generated by the Glimm scheme in spherically symmetric spacetimes in \cite{GroahTemple}, independent of whether the metric itself can be smoothed to $C^{1,1}$. 

\begin{Thm} \label{thm_noRS}\label{loc_inertial_Thm_sph}
Let $\mathcal{M}$ be a spherically symmetric Lorentz manifold with an $L^\infty$ metric connection $\Gamma$ and Riemann curvature tensor bounded in $L^\infty$. Then, for any point $p\in\mathcal{M}$, there exists a coordinate system $y$ within the $C^{1,1}$ atlas which is locally inertial at $p$, in the sense that a representation of the $L^\infty$ equivalence class of the connection $\Gamma^\alpha_{ \beta\gamma}$ in $y$-coordinates vanishes at $p$ and is Lipschitz continuous at the point $p$.
\end{Thm}

In particular, the Lipschitz continuity of $\Gamma^\alpha_{ \beta\gamma}$ at $p$ is necessary and sufficient to remove the Coriolis terms introduced in \cite{ReintjesTemple2}. Theorem \ref{thm_noRS} thus proves that Coriolis terms are removable and that no strong regularity singularities exist in spherically symmetric spacetimes, but it remains open as to whether the metric can always be smoothed to $C^{1,1}$ at points of shock wave interaction.
Thus, the problem of whether (weak) regularity singularities can be created by the Glimm scheme is still an open question.

\section{Preliminaries}   \label{sec_prelim} 

Let $\Gamma$ be an arbitrary connection and ${\rm Riem}(\Gamma)$ its Riemann curvature tensor defined in some open set $\mathcal{U} \subset \mathbb{R}^n$. Assume the components $\Gamma^k_{ij}$ are given $L^{\infty}$ functions in $x$-coordinates, and introduce the components $R^{k}_{\ lij}$ of ${\rm Riem}(\Gamma)$ as distributions on the space $C^{\infty}_0$ of smooth test functions with compact support.  For a smooth connection $\Gamma$, the coefficients of ${\rm Riem}(\Gamma)$ are 
\beq \label{Riemann_prelim} 
{R^k}_{lij}\equiv  Curl(\Gamma)^k_{\ lij}  + [\Gamma_i,\Gamma_j]^k_{l} ,
\eeq
where $\Gamma_i$ denotes the matrix $\Gamma_i\equiv \big( \Gamma^k_{ij} \big)_{k,j=1,...,n}$ and 
\beq \label{curl_plus_commutator}
Curl(\Gamma)^k_{\ lij} \equiv \Gamma^k_{l[j,i]} \equiv \Gamma^k_{lj,i} - \Gamma^k_{li,j} 
\ \ \ \ \ {\rm and} \ \ \ \ \ 
[\Gamma_i,\Gamma_j]^k_{l} \equiv \Gamma^k_{i\sigma } \Gamma^\sigma_{jl} - \Gamma^k_{j\sigma} \Gamma^\sigma_{il} 
\eeq
give the ``curl'' and ``commutator''  terms, respectively. The comma denotes differentiation with respect to $x$. 

Since we are only interested in smoothing $\Gamma$ in a neighborhood of $p$, a local problem, we can without loss of generality assume that $\Gamma^k_{ij}$ are functions of compact support in $\mathcal{U}$. That is, for any neighborhood $\mathcal{U}'$ of $p$ which is compactly contained in $\mathcal{U}$, we can replace $\Gamma$ by $\Gamma'=\varphi\Gamma$ where $\varphi$ is a scalar $C^{\infty}$ function of compact support in $\mathcal{U}$, such that $\varphi(x)=1$ for $x\in\mathcal{U}'$.  Then it is easy to see that both $\Gamma'$ and ${\rm Riem}(\Gamma')$ remain in  $L^{\infty}$, have compact support in $\mathcal{U}$, and agree with $\Gamma$ and ${\rm Riem}(\Gamma)$ in the smaller neighborhood $\mathcal{U}'$.  Thus finding a coordinate system which smooths $\Gamma'$ to $C^{0,1}$ suffices to smooth $\Gamma$ in the neighborhood $\mathcal{U}'$ of $p$.   To see that ${\rm Riem}(\Gamma')$ remains in $L^\infty$, note that the commutator term in \eqref{Riemann_prelim} is in $L^\infty$ since it contains no derivatives, while
\begin{eqnarray} \nonumber
Curl(\Gamma')^k_{\ lij} 
&=& Curl(\Gamma)^k_{\ lij} \varphi  + \Gamma^k_{lj}\varphi_{,i} - \Gamma^k_{li} \varphi_{,j} \ \in L^{\infty}(\mathcal{U}),
\end{eqnarray}
because $Curl(\Gamma)^k_{\ lij}$ and $\Gamma^k_{ij}$ are assumed to be in $L^\infty$ and $\varphi$ in $C^{\infty}$. Thus without loss of generality, we always assume $\Gamma$ has compact support in $\mathcal{U}$, and our analysis can be done over $\mathbb{R}^n$.

Now, for an $L^\infty$ connection $\Gamma^k_{ij}$, the components of ${\rm Riem}(\Gamma)$ are defined as linear functionals in the sense of the theory of distributions. That is,  we can extend the $L^{\infty}$ function $\Gamma$, which has compact support in $\mathcal{U}$, as zero outside of $\mathcal{U}$ to $\mathbb{R}^n$, and define
\begin{eqnarray} \label{curvature_weakform}
 {R}^k_{\ lij}[\psi ]  & \equiv & -\int_{\mathbb{R}^n} \left( \Gamma^k_{lj} \, \psi_{,i} - \Gamma^k_{li} \, \psi_{,j} \right)dx  + \int_{\mathbb{R}^n} [\Gamma_i,\Gamma_j]^k_{l} \,\psi \, dx,
\end{eqnarray}
where $\psi\in C^{\infty}_0(\mathbb{R}^n)$ is any function of compact support in $\mathbb{R}^n$. Similarly, the curl of $\Gamma \in L^\infty(\mathcal{U})$ is defined by
\beq \label{Curl_weak}
Curl(\Gamma)^k_{\ lij}[\psi] \equiv   -\int_{\mathbb{R}^n} \left( \Gamma^k_{lj} \, \psi_{,i} - \Gamma^k_{li} \, \psi_{,j} \right)dx,
\eeq
where again $\psi\in C^{\infty}_0(\mathbb{R}^n)$. Now ${\rm Riem}(\Gamma)$ is bounded in $L^\infty$ if there exists $L^\infty$ functions ${R}^k_{\ lij}$ such that
\beq \nonumber
 R^k_{\ lij}[\psi ] = \int_{\mathcal{U}} {R}^k_{\ lij} \, \psi \, dx = \int_{\mathbb{R}^n} {R}^k_{\ lij} \, \psi \, dx,
\eeq
and in this case $R^k_{\ lij}$ denotes the $L^\infty$ function as well as the distribution.  Thus ${\rm Riem}(\Gamma)=0$ if $R^k_{\ lij}=0$ as an $L^\infty$ function.   
The starting point for this paper is the transformation law for connections
\begin{eqnarray} \label{transfo_connection}
\Gamma^{k}_{ij} 
=   J^k_{\alpha}  \;  \Gamma^{\alpha}_{\beta\gamma}   J^{\beta}_i J^{\gamma}_j             
 +    J^k_{\alpha} \;  \frac{\partial^2 y^\alpha}{\partial x^i \partial x^j},
\end{eqnarray} 
where $J^\alpha_i \equiv \frac{\partial y^{\alpha}}{\partial x^i}$ is the Jacobian of a $C^{1,1}$ coordinate transformation $y\circ x^{-1}$, and $J^i_\alpha$ denotes its inverse.   From \eqref{curvature_weakform} and  \eqref{transfo_connection} one can verify that the class of $L^{\infty}$ connections with $L^{\infty}$ curvature tensors is preserved by the atlas of $C^{1,1}$ coordinate transformation $y\circ x^{-1}$, c.f. \cite{SmollerTemple}, and it shows that the Riemann curvature transforms as a tensor under $C^{1,1}$ coordinate transformations.

\section{Proof of Theorem \ref{thm_equivalence}} \label{sec_geometric_equivalence}

We first give the proof of Theorem \ref{thm_equivalence} assuming Proposition \ref{thm1}, and postpone the proof of Proposition \ref{thm1} until the next section.   Note first that the splitting of $\Gamma^k_{ij}$ into a connection and a $(1,2)$-tensor is consistent with the covariant transformation law (\ref{transfo_connection}), because the difference between two connections is always a tensor, c.f. \cite{HawkingEllis}. That is, assuming $\hat{\Gamma}$ transforms under a coordinate transformation  $y^\alpha\circ x^{-1}$ by the transformation rule of a connection,               
\begin{eqnarray} \label{2} 
\hat{\Gamma}^{\alpha}_{\beta\gamma} =\left\{\hat{\Gamma}^{i}_{jk}J^{j}_{\beta}J^{k}_{\gamma}J^{\alpha}_{i} +J^\alpha_i \frac{\partial^2x^i}{\partial y^{\beta}y^{\gamma}}\right\},
\end{eqnarray} 
and $\tilde{\Gamma}$ transforms by the transformation law of a tensor, 
$$
 \tilde{\Gamma}^{\alpha}_{\beta\gamma}=\tilde{\Gamma}^{i}_{jk}J^{j}_{\beta}J^{k}_{\gamma}J^{\alpha}_{i},
$$
where $J^\alpha_k \equiv \frac{\partial y^\alpha}{\partial x^k}$ is the Jacobian and $J^k_\alpha \equiv \frac{\partial x^k}{\partial y^\alpha}$ its inverse,  it follows that $\Gamma \equiv \tilde{\Gamma} +\hat{\Gamma}$ transforms as a connection, 
\beq \label{2b}
\Gamma^{\alpha}_{\beta\gamma} 
\equiv \tilde{\Gamma}^{\alpha}_{\beta\gamma} + \hat{\Gamma}^{\alpha}_{\beta\gamma} 
=\tilde{\Gamma}^{i}_{jk}J^{j}_{\beta}J^{k}_{\gamma}J^{\alpha}_{i} +\left\{\hat{\Gamma}^{i}_{jk}J^{j}_{\beta}J^{k}_{\gamma}J^{\alpha}_{i} +J^\alpha_i \frac{\partial^2x^i}{\partial y^{\beta}y^{\gamma}}\right\}.
\eeq

To prove the backward implication of Theorem \ref{thm_equivalence} assuming Proposition \ref{thm1}, suppose there exists a splitting $\Gamma^k_{ij}=\tilde{\Gamma}^k_{ij} +\hat{\Gamma}^k_{ij}$ in a neighborhood of $p$ such that $\tilde{\Gamma}^k_{ij}\in C^{0,1}$ is a $(1,2)$-tensor and such that $\hat{\Gamma}^k_{ij}\in L^\infty$ is a connection with ${\rm Riem}(\hat{\Gamma})=0$. By Proposition \ref{thm1}, ${\rm Riem}(\hat{\Gamma})=0$ implies that there exists a coordinate transformation $y\circ x^{-1}$ within the atlas of $C^{1,1}$ transformations such that in $y$-coordinates 
$$ 
\hat{\Gamma}^{\alpha}_{\beta\gamma}=0 
$$ 
in an $L^{\infty}$ almost everywhere sense in a neighborhood of $p$, and hence can be assumed to vanish everywhere.  Thus, by (\ref{2}) and (\ref{2b}), we have in $y$-coordinates that      
$$
 \Gamma^{\alpha}_{\beta\gamma}=\tilde{\Gamma}^{i}_{jk}J^{j}_{\beta}J^{k}_{\gamma}J^{\alpha}_{i} \ \in C^{0,1}.
$$
This establishes the backward implication.

For the forward implication of Theorem \ref{thm_equivalence}, assume there exists a transformation $y\circ x^{-1} \in C^{1,1}(\mathcal{U})$ such that $\Gamma^{\alpha}_{\beta\gamma}\in C^{0,1}(\mathcal{U})$ in $y$-coordinates in some neighborhood $\mathcal{U}$ of $p$. In this case, considering
\begin{eqnarray}\label{3}
\Gamma^{k}_{ij}= \Gamma^{\alpha}_{\beta\gamma} J_{i}^{\beta}J_{j}^{\gamma} J_{\alpha}^{k}+ J_{\alpha}^{k} \frac{\partial^2 y^{\alpha}}{\partial x^{i}\partial x^{j}},
\end{eqnarray} 
we assume without loss of generality that $y \circ x^{-1}$ is the identity outside a compact set in $\mathcal{U}$ containing the support of $\Gamma$. As in the case of $\Gamma$, this can easily be done using smooth cut-off functions.  Thus we assume without loss of generality that $\frac{\partial^2 y^{\alpha}}{\partial x^{i}\partial x^{j}}$ (like $\Gamma$) has compact support in $\mathcal{U}$, and both $\frac{\partial^2 y^{\alpha}}{\partial x^{i}\partial x^{j}}$ and $\Gamma$ extend to $\mathbb{R}^n$ as zero outside of $\mathcal{U}$.  

We now define
$$
\tilde{\Gamma}^k_{ij} \equiv\Gamma^{\alpha}_{\beta\gamma} J_{j}^{\beta} J_{i}^{\gamma} J_{\alpha}^{k} \ \in C^{0,1}(\mathbb{R}^n)
$$
as the Lipschitz continuous tensor part of $\Gamma^k_{ij}$, and 
\begin{eqnarray}\label{4}
\hat{\Gamma}^k_{ij} \equiv J_{\alpha}^{k} \frac{\partial^2 y^{\alpha}}{\partial x^{i}\partial x^{j}}
\end{eqnarray}
as the $L^\infty$ connection part, defined on $\mathbb{R}^n$ with compact support in $\mathcal{U}$.  We now claim the right hand side of (\ref{4}) is flat, satisfying ${\rm Riem}(\hat{\Gamma})=0$ in a neighborhood of $p$, because it is the $y$-coordinate representation of the zero connection in $x$-coordinates. This follows since a curvature tensor in $L^{\infty}$ transform as tensors. To see this explicitly, take the curl of (\ref{4}) in the weak sense (\ref{curvature_weakform}) and observe that the third order (weak) derivatives cancel because
\begin{eqnarray} \nonumber
Curl(\hat{\Gamma})^k_{\ lij}[\psi]  
&=& -\int_{\mathbb{R}^n} J^k_\alpha \left( y^\alpha_{\, ,lj} \psi_{,i} - y^\alpha_{\, ,li} \psi_{,j} \right) dx \cr 
&=& \int_{\mathbb{R}^n} \left( J^k_{\alpha ,i}  y^\alpha_{\, ,lj}  - J^k_{\alpha ,j}  y^\alpha_{\, ,li}  \right) \psi \; dx,
\end{eqnarray}
due to the symmetry in $i$ and $j$, where $\psi\in C^\infty_0(\mathbb{R}^n)$. Thus, the components of the Riemann curvature tensor of (\ref{4}) are in fact given by the $L^\infty$ functions
\beq \nonumber
R^k_{lij} =  J^k_{\alpha ,i}  y^\alpha_{\, ,lj}  - J^k_{\alpha ,j}  y^\alpha_{\, ,li}   
+ \Gamma^k_{i m} \Gamma^m_{jl}  - \Gamma^k_{j m} \Gamma^m_{il}.
\eeq
Using now that $J^\gamma_k J^k_{\alpha,i} = - J^\gamma_{k,i} J^k_\alpha = - y^\gamma_{\, ,ki} J^k_\alpha$, it follows that
\beq \nonumber
J^\gamma_k R^k_{lij} 
= -y^\gamma_{\, ,ki} J^k_\alpha y^\alpha_{,lj}  +y^\gamma_{\, ,kj} J^k_\alpha  y^\alpha_{,li}   
+ J^\gamma_k \Gamma^k_{i \sigma} \Gamma^\sigma_{jl}  - J^\gamma_k \Gamma^k_{j \sigma} \Gamma^\sigma_{il},
\eeq
and substituting (\ref{4}) for the remaining $\Gamma$'s, the above terms mutually cancel to give $J^\gamma_k R^k_{lij} =0$. This establishes the forward implication. 

Finally note that if either side of the equivalence holds we have $Riem(\Gamma) \in L^{\infty}$, because the derivative of the $C^{0,1}$ tensor $\tilde{\Gamma}$ is in $L^\infty$, and this gives \eqref{Riem_bounded_Thm_equiv}. Moreover, equation \eqref{4} implies that the smoothing transformation $y^\alpha \circ x^{-1}$ satisfies \eqref{formulafory}. This completes the proof of Theorem \ref{thm_equivalence}. 
\hfill $\Box$
\vspace{.2cm}

\section{Proof of Proposition \ref{thm1}}  

Assume $\hat{\Gamma}^k_{ij}$ are the components of an $L^{\infty}$ connection defined  in a neighborhood of $p\in\mathcal{M}$ in $x$-coordinates satifying the $L^{\infty}$ bound (\ref{sup}).   For the backward implication of Proposition \ref{thm1},  assume there exists a $C^{1,1}$ transformation $y\circ x^{-1}$ such that             
$\hat{\Gamma}^{\alpha}_{\beta\gamma}=0,$ almost everywhere  in $y$-coordinates.  It follows that ${\rm Riem}(\hat{\Gamma})=0$ in $y$-coordinates in the $L^{\infty}$ sense (\ref{curvature_weakform}), and since the curvature transforms as a tensor under $C^{1,1}$ coordinate transformations, it is ${\rm Riem}(\hat{\Gamma})=0$ in all coordinates of the $C^{1,1}$ atlas.   This establishes the backward implication. 

We now prove the forward implication of Proposition \ref{thm1}, which is the main technical point.   For this, assume $\hat{\Gamma}^k_{ij}(x)$ is an $L^\infty$ connection given in some neighborhood $\mathcal{U}$ in $x$-coordinates, such that ${\rm Riem}(\hat{\Gamma})=0$ in $L^{\infty}$. For the proof we establish a framework in which the classical argument in \cite{Spivak} can be extended to the weaker setting of connections in $L^\infty$. The argument  can be summarized as follows:   The zero curvature condition is used to construct $n$ independent $1$-forms $\omega^\alpha=\omega^\alpha_j dx^j$, ($\alpha=1,...,n$), which are \emph{parallel} in every direction in $\mathcal{U}$, i.e.,  $\nabla_j \omega^\alpha =0$, $j=1,...,n$, where now $\nabla_j$ denotes the covariant derivative for $\hat{\Gamma}$. The parallel condition is then used to construct coordinates $y^\alpha\circ x^{-1}$ 
in which $\hat{\Gamma}$ vanishes.  The difficulty in applying this argument to low regularity $L^{\infty}$ connections is that such connections do not have meaningful restrictions to low dimensional curves and surfaces along which the parallel $1$-forms can be solved for. To remedy this we use a mollification argument.  

The main step in the proof of the forward implication of Proposition \ref{thm1} is stated in Proposition \ref{L1} below. Without loss of generality, we assume from here on that the coordinate neighborhood $\mathcal{U}$ is an $n$-cube, i.e. the direct product of $n$ intervals, $\mathcal{U}=\mathcal{I}_1\times\cdots\times \mathcal{I}_n$, where $\mathcal{I}_k=(a_k,b_k)$ for $a_k<0<b_k$. Again, since our problem is local, there is no loss of generality in assuming that ${\Gamma}$ and $\hat{\Gamma}$ have compact support in $\mathcal{U}$.

\begin{Prop}\label{L1}   
Assume $\hat{\Gamma}$ is a symmetric connection with ${\rm Riem}(\hat{\Gamma})=0$ with  $x$-components $\hat{\Gamma}^k_{ij}(x)$, $x\in \mathcal{U}$, satisfying (\ref{sup}).   Then there exists $n$ linearly independent $1$-forms $\omega^\alpha=\omega^\alpha_{i}dx^i$, $\alpha=1,...,n,$ with components $\omega^\alpha_i$ Lipschitz continuous in $\mathcal{U}$, such that the $1$-forms are parallel almost everywhere, so that
\beq \label{L1parallel}
\left\| \nabla_{j}\omega^{\alpha} \right\|_{L^1(\mathcal{U})} =0,  \ \ \ \ \ \forall j=1,...,n,
\eeq
for $\nabla_j$ the covariant derivative of $\hat{\Gamma}$.  
\end{Prop}      

We now complete the proof of the forward implication of Proposition \ref{thm1} assuming Proposition \ref{L1}. We then give the proof of Proposition \ref{L1} in Section \ref{sec_proof_L1}.  It suffices to construct coordinates $y^{\alpha}$ in which the connection coefficients $\hat{\Gamma}$ vanish.   For this, define
\begin{eqnarray} \label{def_y}
y^\alpha(x^1,...,x^n) 
\equiv  \sum\limits_{i=1}^n \int^{x^i}_0 \omega_i^\alpha(x^1,...,x^{i-1},t,0,...,0)dt,
\end{eqnarray}
where $\omega^{\alpha}$ are the $1$-forms whose existence is guaranteed by Proposition \ref{L1}.  Proposition \ref{L1} implies that \eqref{def_y} defines a $C^{1,1}$ coordinate transformation.  Moreover, equation (\ref{L1parallel}) tells us that $\partial_j \omega^\alpha_i - \hat{\Gamma}^k_{ij} w^\alpha_k$ vanishes in $L^1(\mathcal{U})$, so that the symmetry of the connection $\hat{\Gamma}^k_{ij}= \hat{\Gamma}^k_{ji}$ implies
\beq \label{L1_Cor_techeqn1}
\left\| \partial_{j}\omega^{\alpha}_i - \partial_{i}\omega^{\alpha}_j \right\|_{L^1(\mathcal{U})} =0.
\eeq
Fubini's theorem now implies that there exists a point $x_0$, (which we take without loss of generality to be the origin $x_0=0$ in \eqref{def_y}), such that \eqref{L1_Cor_techeqn1} implies\footnote{Note that if such points $x_0$ did not exist, then one could use this, together with positivity, to integrate up to get a nonzero $L^1$-norm, thereby obtaining a contradiction.}

\beq \label{L1comm_derivatives}
\int_{\Omega_l} \big|\partial_i\omega^{\alpha}_j- \partial_j\omega^{\alpha}_i\big|(x^1,...,x^l,0,...,0) dx^1\cdots dx^l \ =\ 0, \ \ \ \ \ \forall i,j \leq l,
\eeq
for each $l=1,...,n$, where
$
\Omega_l \equiv \mathcal{I}_1 \times\cdots\times \mathcal{I}_l.
$
Differentiating \eqref{def_y}, and integrating over $\mathcal{U}$, we can apply \eqref{L1comm_derivatives}, and thereby commute indices and derivatives on lower dimensional sets in the iterated integrals based at the origin, to obtain, (details are omitted)
\begin{eqnarray}\label{exact2}
\ \ \ \ \ \ \frac{\partial y^\alpha}{\partial x^j} = \omega^\alpha_j,\ \ \ \ \ j,\alpha=1,...,n.
\end{eqnarray} 

Now transforming to $y$-coordinates (\ref{def_y}), the components of $\hat{\Gamma}$ are given by
\begin{eqnarray}\label{exact3}
\hat{\Gamma}^k_{ij}=\frac{\partial^2 y^{\sigma}}{\partial x^i\partial x^j}\frac{\partial x^k}{\partial y^{\sigma}}+\hat{\Gamma}^{\gamma}_{\alpha\beta}\frac{\partial y^{\alpha}}{\partial x^i}\frac{\partial y^{\beta}}{\partial x^j}\frac{\partial x^{k}}{\partial y^{\gamma}}\ \ \ \ \ \text{a.e.} \ \ \text{in} \ \mathcal{U}.
\end{eqnarray}
But by (\ref{exact2}) we have $\frac{\partial y^\alpha}{\partial x^j} = \omega^\alpha_j$, so (\ref{L1parallel}) implies that
\begin{eqnarray} \label{exact3b}
\frac{\partial^2y^{\sigma}}{\partial x^i\partial x^j}=\frac{\partial}{\partial x^i}\omega^{\sigma}_j=\hat{\Gamma}^l_{ij}\omega^{\sigma}_l=\hat{\Gamma}^l_{ij}\frac{\partial y^{\sigma}}{\partial x^l}\ \ \ \ \ \text{a.e.} \ \ \text{in} \ \mathcal{U}.
\end{eqnarray}
Substituting (\ref{exact3b}) into (\ref{exact3}) gives
\begin{eqnarray}\label{exact4}\label{connection_key-argument}
\hat{\Gamma}^k_{ij}=\hat{\Gamma}^k_{ij}+\hat{\Gamma}^{\gamma}_{\alpha\beta}\frac{\partial y^{\alpha}}{\partial x^i}\frac{\partial y^{\beta}}{\partial x^j}\frac{\partial x^{k}}{\partial y^{\gamma}},  \ \ \ \ \ \text{a.e.} \ \ \text{in} \ \mathcal{U},
\end{eqnarray}
and this together with the fact that the Jacobian $\frac{\partial y^{\alpha}}{\partial x^i}$ is non-singular, implies $\hat{\Gamma}^{\gamma}_{\alpha\beta}=0\ \ \text{a.e.} \ \ \text{in} \ \mathcal{U}$. This completes the proof of Proposition \ref{thm1} once we prove Proposition \ref{L1}.  \hfill $\Box$

\section{Proof of Proposition \ref{L1}}  \label{sec_proof_L1}

Assume $\hat{\Gamma}^k_{ij}(x)$ is an $L^\infty$ connection given in some neighborhood $\mathcal{U}$ in $x$-coordinates, such that ${\rm Riem}(\hat{\Gamma})=0$ in the $L^{\infty}$ weak sense.   We construct $n$ linearly independent $1$-forms $\omega^{\alpha}=\omega^{\alpha}_i\,dx^i$ which are Lipschitz continuous and parallel in the sense of (\ref{L1parallel}).   Our strategy is to mollify the connection, and modify the standard argument for constructing parallel $1$-forms when the curvature is zero and the connection is smooth.   The mollified connection, however, has nonzero curvature, so we must keep track of errors in the mollification parameter $\epsilon$ to prove the curvature tends to zero in $L^\infty$ when taking the zero mollification limit at the end.  The basic $L^\infty$ estimates for this are established in Lemmas \ref{Lclaim1} and \ref{Lclaim2} below.   The main technicality in the proof of Proposition \ref{L1} is that the construction requires integrating on lower dimensional surfaces, and the boundary terms arising on these surfaces must also cancel due to zero curvature in the zero mollification limit.   In order to achieve this, we need a peeling property to ensure that the curvature actually vanishes in the zero mollification limit on these lower dimensional sets, c.f. Lemma \ref{Lclaim3} below.  The mollification procedure is also required to apply uniqueness theorems for the ODE's arising from parallel transport.  

To start, let $\phi \in C^\infty_0\big(B_1(0)\big)$ be a standard Gaussian mollifier function with $\int_{\mathbb{R}^n} \phi(x) dx=1$ and we set $\phi_\epsilon(x) \equiv \epsilon^{-n} \phi(\frac{x}{\epsilon})$, so that $\int_{\mathbb{R}^n} \phi_\epsilon(x) dx = 1$ as well. We now consider a standard mollification of $\hat{\Gamma}^k_{ij}(x)$, $x \in \mathbb{R}^n$,             
\begin{eqnarray}\label{mollify2}
(\hat{\Gamma}_{\epsilon})^k_{ij}(x)
= \int_{\mathbb{R}^n}\hat{\Gamma}^{k}_{ij}(\tilde{x})\phi_{\epsilon}(x-\tilde{x})d\tilde{x},
\end{eqnarray}               
where $\hat{\Gamma}$ is assumed without loss of generality to have compact support in $\mathcal{U}$, and is extended as zero outside of $\mathcal{U}$, c.f. \eqref{curvature_weakform} and the preceding discussion. By these definitions, 
$$
\hat{\Gamma}_{\epsilon} \in C^\infty_0(\mathbb{R}^n),
$$ 
with support in a region order epsilon larger than the support of $\hat{\Gamma}$, and $\hat{\Gamma}_{\epsilon}$ converges to $\hat{\Gamma}$ in $L^\infty(\mathbb{R}^n)$ (and hence also in $L^\infty(\mathcal{U})$ and $L^1(\mathcal{U})$) as $\epsilon \rightarrow 0$, and
 $\| \hat{\Gamma}_\epsilon\|_{L^\infty(\mathcal{U})}$ is bounded independently of $\epsilon$,
\beq \label{mollify4}
\| \hat{\Gamma}_\epsilon\|_{L^\infty(\mathbb{R}^n)} 
\leq \|\hat{\Gamma}\|_{L^\infty(\mathcal{U})} \int_{\mathbb{R}^n} \left|\phi_\epsilon(x-\tilde{x}) \right| d\tilde{x}  
\leq \|\hat{\Gamma}\|_{L^{\infty}(\mathcal{U})} .
\eeq
To construct  $1$-forms $\omega^{\alpha}=\omega^{\alpha}_i\,dx^i$, we now establish three lemmas regarding the curvature $\text{\rm Riem}(\hat{\Gamma}_\epsilon)$.

\begin{Lemma} \label{Lclaim1}    
Assume ${\rm Riem}(\hat{\Gamma})$ is bounded in $L^\infty(\mathcal{U})$. Then the mollified curvature satisfies the $\epsilon$-independent bound 
\beq \label{Lclaim1_eqn}
\big\|{\rm Riem}(\hat{\Gamma}_\epsilon)\big\|_{L^\infty(\mathcal{U})} \leq c \big\| \hat{\Gamma} \big\|^2_{L^\infty(\mathcal{U})}  + \big\| {\rm Riem}(\hat{\Gamma})\big\|_{L^\infty(\mathcal{U})},
\eeq
where $c$ is a combinatorial constant depending only on $n$.
\end{Lemma}

\proof Recall that the Riemann curvature tensor can be written as a curl plus a commutator,
\begin{eqnarray}\label{riem0}
{\rm Riem}(\hat{\Gamma}) &\equiv & Curl(\hat{\Gamma})+[\hat{\Gamma},\hat{\Gamma}],
\end{eqnarray}
c.f. (\ref{Riemann_prelim}) - (\ref{curl_plus_commutator}). For the mollified ``curl-part'' of the curvature, observe that
\begin{align} \label{riem0_techeqn0}
Curl_x(\hat{\Gamma}_\epsilon)^k_{\ lij}
&=   \frac{\partial}{\partial x^j} (\hat{\Gamma}_\epsilon)^k_{li}(x) - \frac{\partial}{\partial x^i} (\hat{\Gamma}_\epsilon)^k_{lj}(x) \cr
&= \int_{\mathbb{R}^n} \Big( \hat{\Gamma}^k_{li}(\tilde{x}) \frac{\partial}{\partial x^j} \phi_\epsilon(x-\tilde{x}) - \hat{\Gamma}^k_{lj}(\tilde{x}) \frac{\partial}{\partial x^i} \phi_\epsilon(x-\tilde{x}) \Big) d\tilde{x} \cr
&= -\int_{\mathbb{R}^n} \Big( \hat{\Gamma}^k_{li}(\tilde{x}) \frac{\partial}{\partial \tilde{x}^j} \phi_\epsilon(x-\tilde{x}) - \hat{\Gamma}^k_{lj}(\tilde{x}) \frac{\partial}{\partial \tilde{x}^i} \phi_\epsilon(x-\tilde{x}) \Big) d\tilde{x} \; \cr
&= Curl(\hat{\Gamma})[\phi_\epsilon(x-\cdot)]^k_{\ lij}
\end{align}
which is the weak curl of $\hat{\Gamma}$ acting on the test function $\phi_\epsilon(x-\cdot) \in C^\infty_0(\mathbb{R}^n)$, c.f. \eqref{Curl_weak}. Now, because ${\rm Riem}(\hat{\Gamma})$ is assumed to be in $L^\infty$, there exists $L^\infty$ functions that represent the components of the curl of $\hat{\Gamma}$, since the commutator part in (\ref{Riemann_prelim}) contains no derivatives of $\hat{\Gamma}$. Denoting this $L^\infty$ function by $Curl_{\tilde{x}}(\hat{\Gamma}) \in L^\infty$, the previous equations imply
\beq \label{riem0_techeqn1}
Curl_x(\hat{\Gamma}_\epsilon) = \int_{\mathbb{R}^n} Curl_{\tilde{x}}(\hat{\Gamma}) \phi_\epsilon(x-\tilde{x})\; d\tilde{x}.
\eeq 
Using now the splitting (\ref{riem0}) we write (\ref{riem0_techeqn1}) as
\begin{eqnarray}\nonumber
Curl_x(\hat{\Gamma}_\epsilon)(x) 
&=& -\int_{\mathbb{R}^n} \big([\hat{\Gamma},\hat{\Gamma}] - {\rm Riem}(\hat{\Gamma})\big)(\tilde{x}) \; \phi_\epsilon(x-\tilde{x})d\tilde{x} 
\end{eqnarray}
from which we conclude that
\begin{eqnarray} \nonumber
\big\| Curl(\hat{\Gamma}_\epsilon) \big\|_{L^\infty(\mathbb{R}^n)} 
& \leq &\big\| [\hat{\Gamma},\hat{\Gamma}] \big\|_{L^\infty(\mathbb{R}^n)}  + \big\| {\rm Riem}(\hat{\Gamma})\big\|_{L^\infty(\mathbb{R}^n)} \cr
&\leq & \big\| [\hat{\Gamma},\hat{\Gamma}] \big\|_{L^\infty(\mathcal{U})}  + \big\| {\rm Riem}(\hat{\Gamma})\big\|_{L^\infty(\mathcal{U})} \cr
&\leq &  c\big\| \hat{\Gamma} \big\|^2_{L^\infty(\mathcal{U})}  + \big\| {\rm Riem}(\hat{\Gamma})\big\|_{L^\infty(\mathcal{U})}, 
\end{eqnarray}
for some constant $c$, where we used that $\hat{\Gamma}$ is zero outside of $\mathcal{U}$ to restrict the domain. Thus, from the splitting (\ref{riem0}) for $\hat{\Gamma}_\epsilon$, we find 
\begin{eqnarray}\nonumber
\big\|{\rm Riem}(\hat{\Gamma}_\epsilon)\big\|_{L^\infty(\mathbb{R}^n)} 
&\leq &  c \big\| \hat{\Gamma}_\epsilon \big\|^2_{L^\infty(\mathbb{R}^n)} + \big\|Curl(\hat{\Gamma}_\epsilon)\big\|_{L^\infty(\mathbb{R}^n)}     \cr 
&\leq & 2c \big\| \hat{\Gamma} \big\|^2_{L^\infty(\mathcal{U})}  + \big\| {\rm Riem}(\hat{\Gamma})\big\|_{L^\infty(\mathcal{U})} ,
\end{eqnarray}
which gives the $\epsilon$ independent bound  (\ref{Lclaim1_eqn}) and proves Lemma \ref{Lclaim1}. 
\qed

\begin{Lemma} \label{Lclaim2}
Assume ${\rm Riem}(\hat{\Gamma})$ is bounded in $L^\infty$. Then ${\rm Riem}(\hat{\Gamma}_\epsilon)$ converges to ${\rm Riem}(\hat{\Gamma})$ in $L^\infty(\mathbb{R}^n)$, (and hence, by compact support, both in $L^\infty(\mathcal{U})$ and $L^1(\mathcal{U})$), as $\epsilon\rightarrow 0$.
\end{Lemma}

\proof
To start recall that the mollification of a function in $L^\infty$ converges in $L^\infty$ in the zero mollification limit.    Our assumption is that $\hat{\Gamma}$ and $Riem(\hat{\Gamma})$ are both bounded in $L^{\infty}$, and thus also in $L^1$, since we address a bounded domain $\mathcal{U}$.  The lemma follows under this assumptions because the mollification of the derivative of a function is equal to the derivative of the mollified function, (mollification commutes with differentiation),  and since smooth nonlinear functions of such mollified functions converge to the same function in the zero-mollification limit.  Specifically, we implement this principle as follows: By Christoffel's formula the Riemann tensor is
\begin{eqnarray} \label{Lclaim2_eqn1}
{\rm Riem}(\hat{\Gamma}_\epsilon)^k_{lji} 
= (\hat{\Gamma}_\epsilon)^k_{\ l [j,i]} + [(\hat{\Gamma}_\epsilon)_i,(\hat{\Gamma}_\epsilon)_j]^k_l  ,
\end{eqnarray}
where the curl and commutator parts are defined in \eqref{curl_plus_commutator}. From the definition of the mollification \eqref{mollify2} we find for the curl part that
\begin{eqnarray}\nonumber
(\hat{\Gamma}_\epsilon)^k_{\ l [j,i]} 
&=& \partial_{x^i} (\hat{\Gamma}_\epsilon)^k_{\ lj} -  \partial_{x^j} (\hat{\Gamma}_\epsilon)^k_{\ li} \cr
&=& \int_{\mathbb{R}^n} \Big( \hat{\Gamma}^k_{lj}(\tilde{x}) \partial_{x^i} \phi_{\epsilon}(x-\tilde{x}) - \hat{\Gamma}^k_{li}(\tilde{x}) \partial_{x^i} \phi_{\epsilon}(x-\tilde{x}) \Big) d\tilde{x}   \cr
&\overset{\eqref{riem0_techeqn0}}{=}&   Curl(\hat{\Gamma})[\phi_\epsilon(x-\cdot)]^k_{\ lij}   \cr
&=&  \int_{\mathbb{R}^n} \hat{\Gamma}^k_{l[j,i]}(\tilde{x}) \phi_{\epsilon}(x-\tilde{x})  d\tilde{x},
\end{eqnarray}   
so $(\hat{\Gamma}_\epsilon)^k_{\ l [j,i]}$ is the mollification of $\hat{\Gamma}^k_{l[j,i]}$, and since $\hat{\Gamma}^k_{l[j,i]} \in L^\infty(\mathbb{R}^n)$ it follows that $(\hat{\Gamma}_\epsilon)^k_{\ l [j,i]}$ converges to $\hat{\Gamma}^k_{l[j,i]}$ in $L^\infty$ as $\epsilon \rightarrow 0$. On the other hand, the nonlinear commutator part in \eqref{Lclaim2_eqn1} converges in $L^\infty$ to $[\hat{\Gamma}_i,\hat{\Gamma}_j]^k_l$ as $\epsilon \rightarrow 0$. Namely, by \eqref{curl_plus_commutator} we have
$$
[\hat{\Gamma}_i,\hat{\Gamma}_j]^k_{l} = \hat{\Gamma}^k_{i\sigma } \hat{\Gamma}^\sigma_{jl} - \hat{\Gamma}^k_{j\sigma} \hat{\Gamma}^\sigma_{il} 
$$
and for nonlinear products of the above form we find, say for the first term, that
\begin{align}
&\big\| \hat{\Gamma}^k_{i\sigma} \hat{\Gamma}^\sigma_{jl} - (\hat{\Gamma}_\epsilon)^k_{i\sigma } (\hat{\Gamma}_\epsilon)^\sigma_{jl} \big\|_{L^\infty}     \cr
 & \leq   \big\| \hat{\Gamma}^k_{i\sigma} \big( \hat{\Gamma}^\sigma_{jl} -  (\hat{\Gamma}_\epsilon)^\sigma_{jl} \big) \big\|_{L^\infty}    +     \big\| \big(\hat{\Gamma}^k_{i\sigma}  - (\hat{\Gamma}_\epsilon)^k_{i\sigma } \big) (\hat{\Gamma}_\epsilon)^\sigma_{jl} \big\|_{L^\infty}     \cr
  & \leq   \big\| \hat{\Gamma}^k_{i\sigma}\big\|_{L^\infty}   \big\| \hat{\Gamma}^\sigma_{jl} -  (\hat{\Gamma}_\epsilon)^\sigma_{jl} \big\|_{L^\infty}    +     \big\| \hat{\Gamma}^k_{i\sigma}  - (\hat{\Gamma}_\epsilon)^k_{i\sigma } \big\|_{L^\infty}  \big\| (\hat{\Gamma}_\epsilon)^\sigma_{jl} \big\|_{L^\infty}   ,
\end{align}
where the above norms are taken with respect to $\mathbb{R}^n$. This implies the sought after $L^\infty$ convergence of the commutator term, since $\| \hat{\Gamma}_\epsilon \|_{L^\infty}   \leq \| \hat{\Gamma}\|_{L^\infty}$ by \eqref{mollify4}. Taken on whole, this establishes that ${\rm Riem}(\hat{\Gamma}_\epsilon)$ converges to ${\rm Riem}(\hat{\Gamma})$ in $L^\infty(\mathbb{R}^n)$ as $\epsilon\rightarrow 0$, and since $\hat{\Gamma}$ has compact support this implies convergence in both $L^\infty(\mathcal{U})$ and $L^1(\mathcal{U})$ as $\epsilon \rightarrow 0$. This completes the proof of the Lemma.
\qed \\

The following lemma establishes the $L^1$-{\it peeling property} crucial for assigning initial data consistently in the construction of parallel one-forms (\ref{L1parallel}).

\begin{Lemma} \label{Lclaim3}
Assume ${\rm Riem}(\hat{\Gamma})=0$ in $\mathcal{U} = \mathcal{I}_1\times...\times \mathcal{I}_n$. For every sequence $\epsilon \rightarrow 0$ there exists a subsequence $\epsilon_{k}$ (with $\epsilon_{k} \rightarrow 0$ as $k \rightarrow \infty$) and some point $(\bar{x}^1,..., \bar{x}^n)\in \mathcal{I}_1\times...\times \mathcal{I}_n$ such that the mollified curvature satisfies the $L^1$ \emph{peeling property} at $\bar{x}\equiv(\bar{x}^1,..., \bar{x}^n)$, by which we mean that for each $m=1,...,n$,
\beq \label{peeling1}
\lim_{\epsilon_k \rightarrow 0} \int_{\mathcal{I}_1} ... \int_{\mathcal{I}_m} (R_{\epsilon_k})^k_{\ lij} (x^1,...,x^m,\bar{x}^{m+1},...,\bar{x}^n) dx^1\cdots dx^m =0,
\eeq
that is,
\beq \nonumber
\big\| (R_{\epsilon_k})^k_{\ lij} (\cdot,...,\cdot,\bar{x}^{m+1},...,\bar{x}^n) \big\|_{L^1(\mathcal{I}_1\times...\times \mathcal{I}_m)} \longrightarrow 0  \ \ \ \ \text{as} \ \ \epsilon_k \rightarrow 0.
\eeq
\end{Lemma}

\proof
Define $\tilde{x}\equiv (x^1,...,x^m) \in \mathcal{I}_1\times...\times \mathcal{I}_m$ and $\bar{x}\equiv (x^{m+1},...,x^n) \in \mathcal{I}_{m+1}\times...\times \mathcal{I}_n$.   Fubini's Theorem implies that 
\beq \nonumber
(\bar{R}_\epsilon)^k_{\ lij}(\bar{x}) \equiv \int_{\mathcal{I}_1\times...\times \mathcal{I}_m} (R_{\epsilon})^k_{\ lij}(\tilde{x},\bar{x}) d\tilde{x} 
\eeq
is an integrable function over $\mathcal{I}_{m+1}\times...\times \mathcal{I}_n$. Since, ${\rm Riem}(\hat{\Gamma}_\epsilon)$ converges to zero in $L^1(\mathcal{U})$ by Lemma \ref{Lclaim2}, it follows that $(\bar{R}_\epsilon)^k_{\ lij}$ converges to zero in $L^1(\mathcal{I}_{m+1}\times...\times \mathcal{I}_n)$, namely
\beq \nonumber
\int_{\mathcal{I}_{m+1}\times...\times \mathcal{I}_n} 
(\bar{R}_\epsilon)^k_{\ lij}(\bar{x})d\bar{x} = \int_\mathcal{U} (R_{\epsilon})^k_{\ lij}\; dx \longrightarrow 0, \ \ \ \text{as} \ \ \epsilon \rightarrow 0.
\eeq 
Therefore, there exists a subsequence $\epsilon^m_{k}$ (with $\epsilon^m_{k} \rightarrow 0$ as $k \rightarrow \infty$) and some point $\bar{x} \in \mathcal{I}_{m+1}\times...\times \mathcal{I}_n$ at which $(\bar{R}_{\epsilon^m_{k}})^k_{\ lij}(\cdot)$ converges to zero as $k \rightarrow \infty$. For this point $\bar{x}$, it follows that $(R_{\epsilon^m_{k}})^k_{\ lij}(\cdot,\bar{x})$ converges to $0$ in $L^1(\mathcal{I}_{1}\times...\times \mathcal{I}_m\times\{\bar{x}\})$ as $k \rightarrow \infty$.       

Now, applying this construction with respect to $\mathcal{I}_{1}\times...\times \mathcal{I}_n$, we first find a  point $\bar{x}^n \in \mathcal{I}_n$ together with a subsequence $\epsilon_k^{n-1}$ of $\epsilon$ such that
$$
(R_{\epsilon_k^{n-1}})^k_{\ lij}(\;\cdot\; ,\bar{x}^n)  \longrightarrow 0, \ \ \ \ \ \text{in}\ L^1(\mathcal{I}_1\times...\times\mathcal{I}_{n-1}) ,\ \  \ \text{as} \ \ k \rightarrow \infty.
$$ 
Given this convergence on the $n-1$ sub-cube $\mathcal{I}_1\times...\times\mathcal{I}_{n-1}\times\{\bar{x}^n\}$, we again apply the above construction (but now with respect to the sub-cube) to obtain a point $\bar{x}^{n-1}\in \mathcal{I}_{n-1}$ and a subsequence $\epsilon_k^{n-2}$ of $\epsilon_k^{n-1}$ such that 
$$
(R_{\epsilon_k^{n-2}})^k_{\ lij}(\;\cdot\; ,\bar{x}^{n-1},\bar{x}^n) \longrightarrow 0, \ \ \ \ \ \text{in}\ L^1(\mathcal{I}_1\times...\times\mathcal{I}_{n-2}), \ \  \ \text{as} \ \ k \rightarrow \infty.
$$ 
Continuing, we successively find a subsequence $\epsilon_k$ of $\epsilon$ and a point $(\bar{x}^1,...,\bar{x}^n) \in \mathcal{U}$ at which the peeling property (\ref{peeling1}) holds. This proves Lemma \ref{Lclaim3}. \qed \\

Our goal now is to construct $n$ linearly independent $1$-forms $\omega^\alpha_\epsilon= \left(\omega^\alpha_{\epsilon}\right)_idx^i$, $\alpha=1,...,n$, of the mollified connections $\hat{\Gamma}_{\epsilon}$ by parallel translating in $x$-coordinate directions ${\bf e}_1,...,{\bf e}_n$, one direction at a time, starting with initial data given at a point $\bar{x}$ where the peeling property holds to control the $L^1$-norms of the curvature on the initial data.  The resulting $1$-forms $\omega^\alpha_\epsilon=\left(\omega^\alpha_{\epsilon}\right)_idx^i$,  will not be parallel in every direction because the curvature of the mollified connections is in general nonzero. However, since the Riemann curvature converges to zero in $L^1$ as $\epsilon \rightarrow 0$, one can prove that the $\omega^\alpha_\epsilon$ tend to parallel $1$-forms in the limit $\epsilon \rightarrow 0$, once their convergence in $C^{0,1}$ has been established. Concerning this convergence, the uniform $L^{\infty}$ bound on the curvature alone will imply that the resulting $1$-forms are Lipschitz continuous \emph{uniformly} in $\epsilon$, so that the Arzela-Ascoli Theorem yields a convergent subsequence of the $1$-forms $\omega^\alpha_\epsilon$ that converge to Lipschitz continuous $1$-forms $\omega^1_idx^i,...,\omega^n_idx^i$ as $\epsilon\rightarrow0$. 

To begin the construction of the parallel $1$-forms,  assume a sequence $\epsilon\rightarrow0$ such that the curvature satisfies the peeling property (\ref{peeling1}) at the point $\bar{x}=(\bar{x}^1,...,\bar{x}^n)$.  Assume without loss of generality that $\bar{x}=(0,...,0)$, and $\mathcal{I}_k=(-1,1)\equiv \mathcal{I}$ for each $k=1,...,n$.   We begin with the construction of $1$-forms on the two-surface $\mathcal{I}_1\times\mathcal{I}_2\times\{\bar{x}^3\}\times...\times\{\bar{x}^n\}$ which are parallel in the $x^2$-direction, and then extend to $(x^1,...,x^n) \in \mathcal{U}$ by induction, in the following four steps:          
\vspace{.2cm}

\noindent{\bf Step (i):}  First solve for $1$-forms $\omega_{\epsilon}^\alpha=(\omega_{\epsilon}^\alpha)_jdx^j$, for $\alpha=1,...,n$, parallel along the $x^1$-axis by solving the ODE initial value problem
\begin{eqnarray} \label{eqpar1}\label{loc_inertial_ODE1}
\nabla^\epsilon_1\left(\omega^\alpha_{\epsilon}\right)_j(x^1,0)
&\equiv &\frac{\partial\left(\omega^\alpha_{\epsilon}\right)_j}{\partial x^1}(x^1,0) - \left(\hat{\Gamma}_{\epsilon}\right)^k_{1j}\left(\omega^\alpha_{\epsilon}\right)_k(x^1,0)=0,\\ 
\left(\omega^\alpha_{\epsilon}\right)(0,0)&=&{\bf e}^{\alpha}, \label{eqpar2}
\end{eqnarray}
where we suppress the dependence on $(\bar{x}^3,...,\bar{x}^n)=(0,...,0),$ which are fixed. To ensure linearly independent $1$-forms locally, we choose the initial data for the $1$-forms at the point $(0,0)$ to be the $n$-independent coordinate co-vectors ${\bf e}^{\alpha}\equiv d x^\alpha$. For the construction we keep $\alpha$ fixed and, for ease of notation in Steps (i)-(iv), we write $\omega^\epsilon\equiv \omega^\epsilon_idx^i$ instead of $\omega_\epsilon^\alpha \equiv (\omega_\epsilon^\alpha)_idx^i$.

Taking $t=x^1$, (\ref{eqpar1})-(\ref{eqpar2}) is an initial value problem for an ODE of the form
$$
\dot{u}+A_{\epsilon}u=0,
$$
where $u(t)=\left(\omega^{\epsilon}_1(t,0),...,\omega^{\epsilon}_n(t,0)\right)\in\mathbb{R}^n,$ and $(A_{\epsilon})^k_j(t)=(\hat{\Gamma}_{\epsilon})^k_{1j}(t,0)$ is an $n\times n$ matrix which is smooth and bounded in the $L^\infty$ norm, uniformly in $\epsilon$, by (\ref{mollify4}).   Thus the Picard-Lindel\"off existence theorem for ODE's implies there exists a unique local smooth solution $u(t)=\omega^{\epsilon}(t,0)$.  Moreover, the Gr\"onwall inequality together with the $L^\infty$ bound on $A_{\epsilon}$ implies the resulting $1$-forms $\omega^{\epsilon}(x^1,0)\equiv\omega^{\epsilon}(x^1,\bar{x}^2,...,\bar{x}^n)=\omega^{\epsilon}(x^1,0,...,0)$ are bounded, uniformly in $\epsilon$, which then yields Lipschitz continuity in the $x^1$-direction, uniformly in $\epsilon$. Since we are in a bounded domain with $L^\infty$ bounds on $A_\epsilon$, the above Gr\"onwall estimate implies global existence in the domain of definition of $A_\epsilon$, so $\omega^{\epsilon}(x^1,0)$ is defined for all $x^1\in (-1,1)$.            
\vspace{.2cm}

\noindent{\bf Step (ii):}   Assume $\omega^{\epsilon}(x^1,0)$ from Step (i) is given and defined for $x^1\in (-1,1)$. Now use $\omega^{\epsilon}(x^1,0)$ as initial data to solve for the parallel transport in the $x^2$-direction starting from $x^2=0$, by solving the ODE initial value problem
\begin{eqnarray}\label{eqpar3}\label{loc_inertial_ODE2}
\nabla^\epsilon_2\omega^{\epsilon}_j(x^1,x^2)
&\equiv & \frac{\partial\omega^{\epsilon}_j}{\partial x^2}(x^1,x^2) -\left(\hat{\Gamma}_{\epsilon}\right)^k_{2j}\omega^{\epsilon}_k(x^1,x^2)=0,\ \ \\ \label{eqpar4}
\omega^{\epsilon}(x^1,x^2)&=&\omega^{\epsilon}(x^1,0)\ \ \ \text{at} \ \ x^2=0.
\end{eqnarray}              
For fixed $x^1\in(-1,1)$, taking $t=x^2$,  (\ref{eqpar3})-(\ref{eqpar4}) is an initial value problem for          
\begin{eqnarray} \label{odeinu}
\dot{u}+A_{\epsilon}u=0,
\end{eqnarray} 
with $u(t)=\omega^{\epsilon}(x^1,t)\in\mathbb{R}^n$ and $(A_{\epsilon})^k_j(t)=(\hat{\Gamma}_{\epsilon})^k_{2j}(x^1,t)$ an $n\times n$ matrix which is smooth and bounded in the supnorm uniformly in $\epsilon$, according to (\ref{mollify4}).   The Picard-Lindel\"off theorem implies that there exists a unique smooth solution $\omega^\epsilon(x^1,t)$, and for ease we again assume $\omega^\epsilon(x^1,t)$ to be defined throughout the interval $-1< t< 1$ for each $x^1\in(-1,1)$.   The $\epsilon$-independent supnorm bound $\|A_{\epsilon}\|_{L^\infty}\leq \|\hat{\Gamma}\|_\infty$ on $\mathcal{I}_1 \times \mathcal{I}_2$, together with the Gr\"{o}nwall inequality for (\ref{odeinu}), imply the supnorm bound                    
\begin{eqnarray} \label{bound_1form}
\|\omega^{\epsilon}\|_{L^\infty(\mathcal{I}_1 \times \mathcal{I}_2)}\leq K_0,
\end{eqnarray} 
where we use $K_0$ to denote a universal constant depending only on $\hat{\Gamma}$, independent of $\epsilon$.
Moreover, $\|A_{\epsilon}\|_{L^\infty}\leq \|\hat{\Gamma}\|_\infty$ implies that $\omega^{\epsilon}(x^1,x^2)\equiv\omega^{\epsilon}(x^1,x^2,0,...,0)$ satisfies a Lipschitz bound in the $x^2$-direction, 
\begin{eqnarray}\label{lipbound}
\left\|\frac{\partial\omega^{\epsilon}}{\partial x^2}\right\|_{L^{\infty}(\mathcal{I}_1\times\mathcal{I}_2)}\leq K_0,
\end{eqnarray}  
where $K_0$ again depends on $\hat{\Gamma}$, independent of  $\epsilon$.   Controlling the Lipschitz bound (\ref{lipbound}) in the $x^1$-direction is accomplished in Step (iii). 
\vspace{.2cm}

\noindent{\bf Step (iii):}   To obtain Lipschitz continuity of $\omega^{\epsilon}(x^1,x^2)$ in the $x^1$-direction, uniformly in $\epsilon$ and $x^2$, we estimate the change of $u\equiv \nabla^\epsilon_1\omega^{\epsilon}$ in $x^2$-direction, starting from $x^2=0$ where $u\equiv \nabla^\epsilon_1\omega^{\epsilon}(x^1,0)=0$ by construction.  By the definition of curvature, we can write 
\begin{eqnarray}\label{eqpar7}
\nabla^\epsilon_2\left[\omega^{\epsilon}_{k;1}\right]
&=&\nabla^\epsilon_1\nabla^\epsilon_2\omega^{\epsilon}_k +\left(R_{\epsilon}\right)^{\sigma}_{k21}\omega^{\epsilon}_{\sigma},
\end{eqnarray}
so the definition of covariant derivative gives 
\begin{eqnarray}\nonumber
\nabla^\epsilon_1\nabla^\epsilon_2\omega^{\epsilon}_k =\frac{\partial}{\partial x^1} \left[\omega^{\epsilon}_{k;2}\right] -(\hat{\Gamma}_\epsilon)^{\sigma}_{1k}\omega_{\sigma;2}- (\hat{\Gamma}_\epsilon)^{\sigma}_{12}\omega^{\epsilon}_{k;\sigma}\ .
\end{eqnarray}
Substituting this into (\ref{eqpar7}), using $\omega_{\sigma;2}=0$, we find that (\ref{eqpar7}) is equivalent to
\begin{eqnarray}\label{eqpar9}
\nabla^\epsilon_2\left[\omega^{\epsilon}_{k;1}\right] +(\hat{\Gamma}_\epsilon)^{\sigma}_{12}\omega^{\epsilon}_{k;\sigma} -\left(R_{\epsilon}\right)^{\sigma}_{k21}\omega^{\epsilon}_{\sigma} =0.
\end{eqnarray}
On the other hand,  the definition of $\nabla^\epsilon_2$ gives
\begin{eqnarray}\label{eqpar10}
\nabla^\epsilon_2\left[\omega^{\epsilon}_{k;1}\right]=\frac{\partial}{\partial x^2}\left[\omega^{\epsilon}_{k;1}\right] -(\hat{\Gamma}_\epsilon)^{\sigma}_{2k}\omega^{\epsilon}_{\sigma;1} -(\hat{\Gamma}_\epsilon)^{\sigma}_{21}\omega^{\epsilon}_{k;\sigma}\ .
\end{eqnarray}
Substituting (\ref{eqpar10}) into (\ref{eqpar9}), a cancellation gives the ODE for $u\equiv \omega^{\epsilon}_{k;1}$,
\begin{eqnarray} \label{eqpar11}
\frac{\partial}{\partial x^2}\left[\omega^{\epsilon}_{k;1}\right] -(\hat{\Gamma}_\epsilon)^{\sigma}_{2k}\omega^{\epsilon}_{\sigma;1} -\left(R_{\epsilon}\right)^{\sigma}_{k21}\omega^{\epsilon}_{\sigma}=0.
\end{eqnarray}
Thus, for fixed $x^1$, letting $t=x^2$ and
$
u_k(t)\equiv \omega^{\epsilon}_{k;1}(x^1,t),\ \ k=1,...,n,
$
the $x^2$-directional change of $u$ is determined by the system of ODE's
 \begin{eqnarray}\label{eqpar12}
\dot{u}+A_{\epsilon}u+B_{\epsilon}=0,
\end{eqnarray}
where $u=(u_1,...,u_n)$, and the $n\times n$-matrix $A_{\epsilon}$ as well as the $n$-vector $B_{\epsilon}$ are
\beq \nonumber
\left(A_{\epsilon}\right)^{\sigma}_k=-(\hat{\Gamma}_{\epsilon})^{\sigma}_{2k}
\hspace{.6cm} \text{and} \hspace{.6cm}
\left(B_{\epsilon}\right)_k= -\left(R_{\epsilon}\right)^{\sigma}_{k21}\omega^{\epsilon}_{\sigma}.
\eeq

In addition, we have by Lemma \ref{Lclaim1} and (\ref{bound_1form}) that
\beq \label{Bbound1}
\|B_{\epsilon}\|_{L^\infty}\leq \|R_{\epsilon}\|_{L^\infty}\, \|\omega^{\epsilon}_{\sigma}\|_{L^\infty}\leq K_0 ,
\eeq
for some constant $K_0$ independent of $\epsilon$, and the $L^\infty$-norm is taken on $\mathcal{I}_1\times\mathcal{I}_2$. Applying the Gr\"onwall inequality in (\ref{eqpar12}), using the bound on $A_{\epsilon}$, we obtain
\beq \label{Bbound}
|\omega^{\epsilon}_{k;1}|(t) \leq  K_0\int_0^t|B_{\epsilon}|dt, 
\eeq
for $K_0>0$ independent of $\epsilon$. Estimate (\ref{Bbound}) and (\ref{Bbound1}) together with the definition of the covariant derivative, $\omega^{\epsilon}_{i;1}=\omega^{\epsilon}_{i,1}-\left(\hat{\Gamma}_{\epsilon}\right)^{\sigma}_{i1} \omega^{\epsilon}_{\sigma}$,
implies the supnorm of the derivative $\frac{\partial}{\partial x^1}\omega^{\epsilon}$ is bounded uniformly in $\epsilon$ by
\beq \label{Bbound_partialder}
\Big|\frac{\partial\omega^{\epsilon}_{k}}{\partial x^1}\Big|(t) \leq  K\int_0^t|B_{\epsilon}|dt + \big| (\hat{\Gamma}_\epsilon)^\sigma_{k1} w^\epsilon_\sigma \big|\leq K_0.
\eeq 
for some generic constant $K_0$ independent of $\epsilon$.  We conclude that the components
$\omega^{\epsilon}_i(x^1,x^2,0,...,0)$ 
are Lipschitz continuous in $(x^1,x^2)$, uniformly in $\epsilon$.    
\vspace{.2cm}

\noindent {\bf Step (iv):}  In the final step, we use induction to extend the construction of $\omega^{\epsilon}$ , and obtain the Lipschitz estimate corresponding to (\ref{Bbound_partialder}) in $n$-dimensions.   To implement the induction step $m-1$ to $m$, with $m\leq n$, requires controlling $m-1$ commutators of covariant derivatives.  The step $m=3$ is essentially different from $m=2$ because it is at this step that, for example, $\nabla_1\omega$ does not vanish on the initial data surface $\mathcal{I}_1\times\mathcal{I}_2$.  This is the obstacle to constructing locally inertial frames for $n\geq3$ in the next section.   

For the induction assumption, let $\omega^{\epsilon}(x^1,...,x^{m-1},0,...,0)$ be the $1$-form in $C^\infty(\mathcal{I}_1\times...\times \mathcal{I}_{m-1})$ which generalizes the construction in Steps (i) - (ii) as follows: We assume the parallel transport condition,
\beq \label{IndAss2}
\nabla^\epsilon_k \omega^\epsilon(x^1,...,x^k,0,...,0) =0, \ \ \ \ \ \forall k\leq m-1,
\eeq 
and we assume the Lipschitz norm of $\omega^{\epsilon}$ to be bounded uniformly in $\epsilon$ analogously to (\ref{Bbound}). That is, for each $l\leq m-1$ we assume 
$$
\big\|\omega^\epsilon\big\|_{L^\infty(\Omega_{l})} \leq K_0,
$$
where again $K_0$ denotes a constant $K_0$ depending on $\hat{\Gamma}$,  independent of $\epsilon$, and we assume that
\small
\beq \label{IndAss1}
|{\omega}^\epsilon_{k;j}| (x^1,...,x^{m-1},0,...,0) \leq  K_0 \sum\limits_{l=1}^{m-1} \left| \int^{x^l}_{0}\big|\left(R_{\epsilon}\right)^{\sigma}_{klj}\omega^{\epsilon}_{\sigma} \big|(x^1,...,x^{l-1},t,0,...,0) \: dt \right|. 
\eeq
\normalsize
Note that (\ref{IndAss1}) together with the curvature bound from Lemma \ref{Lclaim1} imply the $\epsilon$-independent bound 
\beq \nonumber 
\|\omega^\epsilon\|_{C^{0,1}(\Omega_{m-1})} \leq K_0,
\eeq
where $\Omega_{l} \equiv \mathcal{I}_1\times...\times \mathcal{I}_{l} \times \{0\} \times . . . \times \{0\}$ for $l=1,...,n$ and
$$
\|\omega^\epsilon\|_{C^{0,1}(\Omega_{l})} \equiv \big\|\omega^\epsilon\big\|_{L^\infty(\Omega_{l})} + \sum\limits_{l=1}^{m-1} \big\|\partial_l \omega^\epsilon\big\|_{L^\infty(\Omega_{l})}.
$$

The induction step now is to prove that there exists a $1$-form ${\omega}^\epsilon \in C^\infty(\Omega_m)$ which agrees with $\omega^\epsilon$ when $x^m=0$ and satisfies the Lipschitz bound (\ref{IndAss1}) on $\Omega_m$ for some constant $K_0>0$ independent of $\epsilon$, such that for each $k\leq m$ the parallel transport condition (\ref{IndAss2}) holds. (For ease, we assume that $\mathcal{I}_l= (-1,1)$ for each $l=1,...,n$.)  As in Step (ii), we extend $\omega^{\epsilon}$
from $\Omega_{m-1}$ to $\Omega_m$ by solving the ODE for parallel transport in $x^m$-direction,
\begin{eqnarray} \label{ODE_m_parallel}
\nabla^\epsilon_m \hat{\omega}^\epsilon(x^1...,x^m,0,...,0) = 0
\end{eqnarray}
for fixed $x^1,...,x^{m-1}$ and with  initial data
\begin{eqnarray}
\hat{\omega}^\epsilon (x^1,...,x^{m-1},0,...,0) = \omega^\epsilon(x^1,...,x^{m-1},0,...,0) .\nonumber
\end{eqnarray}
We denote the solution of (\ref{ODE_m_parallel}) again by $\omega^\epsilon\equiv \hat{\omega}^\epsilon$. Analogous to Step (ii), ${\omega}^\epsilon \in C^\infty(\Omega_{m})$ and the parallel condition (\ref{IndAss2}) is satisfied by construction for each $k\leq m$. Moreover, the Gr\"onwall inequality implies that ${\omega}^\epsilon$ is sup-norm bounded over $\Omega_{m}$ and by (\ref{mollify4}) this bound is independent of $\epsilon$.   The $\epsilon$-independent bound on $\|\partial_m \omega^\epsilon\|_{L^\infty(\Omega_m)}$ now follows from (\ref{ODE_m_parallel}).

It remains to prove $\epsilon$-independent bounds on $\|\partial_j \omega^\epsilon\|_{L^\infty(\Omega_m)}$ for each $j<m$ to prove the Lipschitz bound analogous to (\ref{IndAss1}) on $\Omega_m$. For this we prove the following Lemma.

\begin{Lemma} \label{peeling_lemma_step(iv)}
The $1$-forms solving (\ref{ODE_m_parallel}) satisfy
\beq \label{Induction_step_estimate}
|{\omega}^\epsilon_{k;j}| (x^1,...,x^m,0,...,0) 
\leq    K_0 \sum\limits_{l=1}^m  \int_0^{x^l} \big|(R_\epsilon)^\sigma_{klj} \: {\omega}^\epsilon_\sigma \big|(x^1,...,x^{l-1},t,0,...,0) \: dt  ,
\eeq
for some constant $K_0>0$ depending only on $\hat{\Gamma}$, independent of $\epsilon$.
\end{Lemma}

\proof 
We proceed similarly to Step (iii) and use the definition of the curvature tensor to write for each $j<m$
\beq \label{ODE_ind_step1}
\nabla^\epsilon_m\nabla^\epsilon_j {\omega}^\epsilon_k = \nabla^\epsilon_j \nabla^\epsilon_m{\omega}^\epsilon_k + (R_\epsilon)^\sigma_{kmj} {\omega}^\epsilon_\sigma.
\eeq
Computing the components of the covariant derivatives in (\ref{ODE_ind_step1}) in terms of their connection coefficients, using that ${\omega}^\epsilon_{j;m}=0$ for all $j=1,..,n$, we find that
\begin{eqnarray} \nonumber
\nabla^\epsilon_j \nabla^\epsilon_m{\omega}^\epsilon_k 
\ &=&\ \ \partial_j [{\omega}^\epsilon_{k;m}]  - (\hat{\Gamma}_\epsilon)^\sigma_{jk} [{\omega}^\epsilon_{\sigma;m}] - (\hat{\Gamma}_\epsilon)^\sigma_{jm} [{\omega}^\epsilon_{k;\sigma}] 
\ =\ \ - (\hat{\Gamma}_\epsilon)^\sigma_{jm} [{\omega}^\epsilon_{k;\sigma}]
\end{eqnarray}
and
\begin{eqnarray} \nonumber 
\nabla^\epsilon_m \nabla^\epsilon_j {\omega}^\epsilon_k 
&=& \partial_m [{\omega}^\epsilon_{k;j}] - (\hat{\Gamma}_\epsilon)^\sigma_{mk} [{\omega}^\epsilon_{\sigma;j}] - (\hat{\Gamma}_\epsilon)^\sigma_{mj} [{\omega}^\epsilon_{k;\sigma}].
\end{eqnarray}
Substituting the previous two identities into (\ref{ODE_ind_step1}), we find that (\ref{ODE_ind_step1}) is equivalent to the system of ODE's
\beq \label{ODE_ind_step}
\partial_m [{\omega}^\epsilon_{k;j}] - (\hat{\Gamma}_\epsilon)^\sigma_{mk} [{\omega}^\epsilon_{\sigma;j}] - (R_\epsilon)^\sigma_{kmj} {\omega}^\epsilon_\sigma =0.
\eeq
Applying the Gr\"onwall inequality to the ODE (\ref{ODE_ind_step}) leads to the estimate
\begin{eqnarray} \label{Induction_step_estimate_again}
|{\omega}^\epsilon_{k;j}| (x^1,...,x^m,0,...,0) 
&\leq &    K_0 \int_{0}^{x^m} |(R_\epsilon)^\sigma_{kmj} {\omega}^\epsilon_\sigma |(x^1,...,x^{m-1},t,0,...,0) dt \cr & & \ \ \ \ \ + |\omega^\epsilon_{k;j}| (x^1,...,x^{m-1},0,...,0),
\end{eqnarray}
where $K_0>0$ is independent of $\epsilon$ because of (\ref{mollify4}).\footnote{The difference between the Gr\"onwall estimate in (\ref{Induction_step_estimate_again}) and the one in Step (iii) is the presence of the second term on the right hand side which is due to the initial data $\omega^{\epsilon}$ being not parallel for $\epsilon > 0$ and $j\geq 2$.}   Using the induction assumption (\ref{IndAss1}) to replace the initial data term $|\omega^\epsilon_{k;j}| (x^1,...,x^{m-1},0,...,0)$ on the right hand side of (\ref{Induction_step_estimate_again}) gives us the sought after estimate (\ref{Induction_step_estimate}).
\qed \\

The $\epsilon$-independent Lipschitz bound for ${\omega}^{\epsilon}$ on $\mathcal{U} =\mathcal{I}_1\times...\times \mathcal{I}_{n}$ now follows directly from (\ref{Induction_step_estimate}). Namely, $\|(R_\epsilon)^\sigma_{kmj} {\omega}^\epsilon_\sigma\|_{L^\infty} \leq K_0 \|\hat{\Gamma}\|^2_{L^\infty}$ according to Lemma \ref{Lclaim1} and the boundedness of $\|{\omega}^\epsilon_\sigma\|_{L^\infty}$ derived above. We conclude 
\beq \label{final_Lipschitz_estimate}
\big\| w^\epsilon \big\|_{C^{0,1}(\mathcal{U})} \equiv \big\| w^\epsilon \big\|_{L^\infty(\mathcal{U})} \; +\; \sum_{j=1}^n \left\| \frac{\partial w^\epsilon}{\partial x^j} \right\|_{L^\infty(\mathcal{U})} \ \leq \ K_0,
\eeq
for some positive constant $K_0$ depending only on $\hat{\Gamma}$, independent of $\epsilon$. This completes the induction step and proves that the $1$-forms $\omega^{\epsilon}(x^1,...,x^n)$ are Lipschitz continuous, uniformly in~$\epsilon$. This completes Step (iv).  
\vspace{.2cm} 

To summarize, in Steps (i) - (iv) we constructed $n$ families of smooth $1$-forms $(w_\epsilon)^\alpha_i dx^i$, (with $\alpha = 1,...,n$), such that each component satisfies the uniform Lipschitz bound (\ref{final_Lipschitz_estimate}). Thus, for each $\alpha = 1,...,n$, the Arzela-Ascoli Theorem yields a subsequence of the $1$-forms $(w_\epsilon)^\alpha_i dx^i$ that converges uniformly to a Lipschitz continuous $1$-form $(w_\epsilon)^\alpha_i \ \longrightarrow \ \omega^\alpha_i$ as $\epsilon\rightarrow0.$
Since for each $\alpha = 1,...,n$ the initial data in Step (i) was chosen such that each $1$-form $(w_\epsilon)^\alpha_i dx^i$ agrees with the unit co-vector ${\bf e}^\alpha_kdx^k = dx^\alpha$ at the point $(\bar{x}^1,...,\bar{x}^n)=(0,...,0)$ for any $\epsilon >0$, the limit $1$-form $\omega^\alpha_i dx^i$ is identical to $dx^\alpha$ at $(\bar{x}^1,...,\bar{x}^n)$ as well. Thus, the $1$-forms $(w_\epsilon)^\alpha_i dx^i$ are linearly independent and linear independence throughout $\mathcal{U}$ now follows from the uniqueness of solutions of ODE's, c.f. (\ref{eqpar2}).    

To complete the proof of Proposition \ref{L1}, it remains to prove that the limit $1$-forms are parallel in every direction with respect to $\hat{\Gamma}$ in the $L^1$ sense of (\ref{L1parallel}).  For this, integrate the ODE estimate (\ref{Induction_step_estimate}) for $m=n$ over $\mathcal{U} =\mathcal{I}_1\times ...\times \mathcal{I}_n \equiv \Omega_n$, to get
\begin{eqnarray} \label{Ind_step_estimate_2}
\|{\omega}^\epsilon_{k;j}\|_{L^1(\mathcal{U})} 
&\leq&  K_0 \sum_{l=1}^n  \big\|(R_\epsilon)^\sigma_{klj}\big\|_{L^1(\Omega_l)} \big\| \omega^\epsilon_\sigma \big\|_{L^{\infty}(\Omega_n)}
\end{eqnarray}
where $\Omega_l \equiv \mathcal{I}_1\times ...\times \mathcal{I}_l \times \{0\} \times ... \times\{0\} \subset \mathcal{U}$ for $l=1,...,n$ and $K_0>0$ a universal constant independent of $\epsilon$. Since $\omega^\epsilon$ is bounded in $L^\infty(\Omega_n)$ independent of $\epsilon$, the  $L^1$-peeling property of the curvature (\ref{peeling1}) now implies that the right hand side of (\ref{Ind_step_estimate_2}) converges to zero for some subsequence $\epsilon_k\rightarrow 0$. Thus each of the $1$-forms $\omega^\alpha$ is parallel in $L^1(\mathcal{U})$ in every direction, as claimed in (\ref{L1parallel}). This completes the proof of Proposition \ref{L1}. \hfill $\Box$

\section{A Construction of Locally Inertial Frames}

We begin by giving the definition of locally inertial coordinates for $L^{\infty}$ connections in $n$-dimensions: 

\begin{Def}\label{Deflipcont}
Let $\Gamma$ be an $L^{\infty}$ connection.  We say a coordinate system $y$ is locally inertial for $\Gamma$ at $p$ if the components satisfy
\begin{eqnarray}\label{locallyinertial}
&\left|\Gamma^\alpha_{\beta\gamma}(y)\right|\leq K|y-y(p)|\ \ a.e.,
\end{eqnarray}
for some constant $K$ independent of $y$.  We say $\Gamma$ is locally inertial at $p$ if there exists a locally inertial coordinate system at $p$. 
\end{Def}   

Condition (\ref{locallyinertial}) is equivalent to the existence of an $L^{\infty}$ representation of the components $\Gamma^\alpha_{\beta\gamma}(y)$ such that  (\ref{locallyinertial}) holds in the pointwise everywhere sense, and $\Gamma^\alpha_{\beta\gamma}(y(p))=0$.
In this section we use the coordinate construction of Section \ref{sec_proof_L1} to prove that locally inertial coordinates exist for $L^\infty$ connections in $2$-dimensional manifolds when the Riemann curvature tensor of the connection is assumed bounded in $L^{\infty}$.   Building on this construction in $2$-dimensions, we prove that locally inertial frames always exist in $4$-dimensional spherically symmetric spacetimes with Lipschitz continuous metric.
Thus in particular,  it is sufficient to apply to the GR shock wave solutions generated by the Glimm method, \cite{GroahTemple}.    Interestingly, this argument does not extend to three or more dimensions essentially because the induction step in (iv) of the proof of Proposition \ref{L1} at $n>2$ differs from the $n=2$ step by boundary terms arising from Gronwall estimate (\ref{Induction_step_estimate_again}), and these terms would not vanish in the zero mollification limit when the analogue of the peeling property was used for nonzero curvature.   To construct locally inertial coordinates in $2$-dimensional spacetimes, we construct $1$-forms as in Steps (i) and (ii) of Proposition \ref{L1}, (the case ${\rm Riem}(\Gamma)=0$), and define coordinates $y^\alpha$ by integrating over these $1$-forms.   These $1$-forms are not in general parallel, but as a consequence of the $L^\infty$ curvature bound, we prove the $1$-forms are parallel within error of order $O(|x|)$ when curvature is non-zero. This then implies that the connection is order $O(|y|)$ in coordinates $y^\alpha$, the condition that $y^\alpha$ be locally inertial.  Theorem \ref{thm_noRS} of the introduction follows from Proposition \ref{loc_inertial_Thm_2D} and \ref{loc_inertial_Thm_sph} of this section.

\subsection{Locally Inertial Frames in $2$-Dimensions}\label{susec8.1}

The goal of this section is to prove the following theorem:
\begin{Prop} \label{loc_inertial_Thm_2D}
Assume $\mathcal{M}$ is a two-dimensional manifold endowed with a symmetric $L^\infty$-connection with  Riemann curvature tensor bounded in $L^{\infty}$, and let $p\in\mathcal{M}$.   Then there exists locally inertial coordinates  at $p$ within the $C^{1,1}$ atlas.
\end{Prop}

To prove Proposition \ref{loc_inertial_Thm_2D}, assume $n=2$ in the constructions of Steps (i) and (iii) of Proposition \ref{L1}.
Then for each $\alpha=1,2$, we have a subsequence of the family of $1$-forms $(\omega_\epsilon)^\alpha$ which converges to a Lipschitz continuous $1$-form $\omega^\alpha$ as $\epsilon \rightarrow 0$. By (\ref{loc_inertial_ODE2}), each $\omega^\alpha$ is parallel in the $x^2$-direction in the $L^1$ sense 
\beq \label{L1_parallel_loc_int}
\|\nabla_2 w^\alpha\|_{L^1}=0 .
\eeq
However, in contrast to Section \ref{sec_proof_L1}, we cannot expect $\omega^\alpha$ to be parallel in the $x^1$-direction when $Riem(\Gamma)\neq0$. However, as a result of the $L^\infty$ curvature bound, the $\omega^\alpha$ are approximately parallel in the sense of the following lemma.
 
\begin{Lemma} \label{loc_inertial_Lemma1}
The $1$-forms $\omega^\alpha$, obtained from the zero-mollification limit of (\ref{loc_inertial_ODE2}), satisfy
\beq \label{loc_inertial_Lemma1_eqn} 
\Big| \frac{\partial\omega^\alpha_j}{\partial x^i} - \Gamma^k_{ij} \omega^\alpha_k \Big|(x^1,x^2) \leq K_0 \big(|x^1|+|x^2|\big)\equiv O(x)
\eeq
almost everywhere, where $K_0>0$ is some constant depending only on $\|\Gamma\|_{L^\infty}$ and $\|{\rm Riem}(\Gamma)\|_{L^\infty}$.
\end{Lemma}

\proof
Equation (\ref{L1_parallel_loc_int}) immediately implies (\ref{loc_inertial_Lemma1_eqn}) for~$i=2$ because the right hand side of(\ref{L1_parallel_loc_int}) vanishes when $i=2$.   It remains only to verify (\ref{loc_inertial_Lemma1_eqn}) for~$i=1$.   For the case $i=1$, observe that
 the computation (\ref{eqpar7}) - (\ref{eqpar11}) of Step (iii) in Section \ref{sec_proof_L1}, again gives that the ODE (\ref{loc_inertial_ODE2}) implies
\begin{eqnarray} \label{2dim5}
\frac{\partial u_k}{\partial x^2} &=& (\Gamma_\epsilon)^{\sigma}_{2k}u_\sigma + (R_\epsilon)^\sigma_{k12} (\omega_\epsilon)^\alpha_\sigma
\end{eqnarray}
for $u_k \equiv  \nabla_1 (\omega_\epsilon)^\alpha_k$ and where $(R_\epsilon)^\sigma_{kij}$ denotes the components of ${\rm Riem}(\Gamma_\epsilon)$.
Applying the Gr\"onwall inequality to (\ref{2dim5}) and the fact that $\nabla_1\omega^\alpha_k(x^1,0)=0$ by (\ref{loc_inertial_ODE1}), we obtain
\beq \label{intest}
|\nabla_1(\omega_\epsilon)^\alpha_k|(x^1,x^2) \leq  K_0 \int_0^{x^2}|(R_\epsilon)^\sigma_{k12} (\omega_\epsilon)^\alpha_\sigma|(x^1,t)dt, 
\eeq
where here $K_0>0$ always denotes a generic constant depending on $\|\Gamma\|_{L^\infty}$  and $\|{\rm Riem}(\Gamma)\|_{L^\infty}$, but independent of $\epsilon$. Using that (\ref{loc_inertial_ODE2}) implies $\|\omega^\alpha_\sigma\|_{L^\infty}< K_0 \|\Gamma\|_{L^\infty}$ together with the curvature bound (\ref{Lclaim1_eqn}), we obtain from (\ref{intest}) the further estimate
\beq \label{2dim5_Groenwall}
|\nabla_1(\omega_\epsilon)^\alpha_k|(x^1,x^2) \ \leq\  K_0 \|\Gamma\|_{L^\infty} \max_{\sigma=1,2}\big\|R^\sigma_{k12}\big\|_{L^\infty} \: \big|x^2\big| \ \equiv \ K_0 \big|x^2\big|.
\eeq
Now $\Gamma_\epsilon$ converges in $L^1(\mathcal{U})$ as $\epsilon \rightarrow 0$, so there exists a subsequence converging pointwise almost everywhere. From this pointwise convergence and the fact that $\frac{\partial}{\partial x^1}(\omega_\epsilon)^\alpha$ converges in $L^\infty(\mathcal{U})$, we conclude that 
\beq \label{2dim5_Groenwall_again}
|\nabla_1\omega^\alpha_k|(x^1,x^2) \ \leq \ K_0 \big|x^2\big|\ \ \ \ a.e., 
\eeq
which is the sought after error estimate (\ref{loc_inertial_Lemma1_eqn}) for $i=1$.
\qed \\

To prove Proposition \ref{loc_inertial_Thm_2D}, we define for each $\alpha=1,2$ the coordinates $y^{\alpha}$ on $\mathcal{U}$ by
\beq \label{2dim1}
y^{\alpha}(x^1,x^2) \equiv \int_0^{x^1}\omega^{\alpha}_1(s,x^2)ds+\int_0^{x^2}\omega^{\alpha}_2(0,s)ds,
\eeq
and complete the proof by showing $y^{\alpha}$ are locally inertial at $p$.  By the definition of $y^{\alpha}$ we have
\beq\label{whatwewant1}
\frac{\partial y^\alpha}{\partial x^1}=\omega^{\alpha}_1 
\hspace{1cm} \text{and} \hspace{1cm} 
\frac{\partial}{\partial x^j} \frac{\partial y^\alpha}{\partial x^1}=\frac{\partial\omega^\alpha_1}{\partial x^j},
\eeq
which, for $i=1$, is the identity that leads to the cancellation in (\ref{connection_key-argument}). However, we cannot obtain these identities for the $x^2$-derivative because the $1$-forms $\omega^\alpha$ are no longer parallel in the $x^1$-direction.   The following approximate identities are sufficient for the existence of locally inertial frames. 

\begin{Lemma} \label{loc_inertial_Lemma2}
The coordinates $y^\alpha$ defined in (\ref{2dim1}) satisfy for $i,j=1,2$
\begin{eqnarray} 
\left| \frac{\partial y^\alpha}{\partial x^i} -  w^\alpha_i \right|(x^1,x^2) &\leq &  K_0 \big(|x^1|+|x^2|\big), \label{loc_inertial_Lemma2_eqn1}  \\
\left| \frac{\partial^2y^\alpha}{\partial x^j\partial x^i} - \frac{\partial\omega^\alpha_i}{\partial x^j} \right|(x^1,x^2) &\leq &  K_0 \big(|x^1|+|x^2|\big) \ \ \ \ \ \text{a.e.}, \label{loc_inertial_Lemma2_eqn2} 
\end{eqnarray}
where $K_0>0$ is some constant depending only on $\|\Gamma\|_{L^\infty}$ and $\|{\rm Riem}(\Gamma)\|_{L^\infty}$.
\end{Lemma}

\proof
The case $i=1$ follows directly from (\ref{whatwewant1}).   For the case $i=2$, differentiate (\ref{2dim1}) in the $x^2$ direction to get
\beq \nonumber
\frac{\partial y^{\alpha}}{\partial x^2}(x^1,x^2) 
= \int_0^{x^1}\frac{\partial\omega^\alpha_1}{\partial{x^2}}(s,x^2)ds+\omega^{\alpha}_2(0,x^2) .
\eeq
Using that $\frac{\partial(\omega_\epsilon)^\alpha_1}{\partial{x^2}}$ converges to $\frac{\partial\omega^\alpha_1}{\partial{x^2}}$ in $L^1(\mathcal{U})$ as $\epsilon \rightarrow 0$, the dominated convergence theorem implies that
\beq \label{2dim2}
\frac{\partial y^{\alpha}}{\partial x^2}(x^1,x^2)  =  \lim_{\epsilon\rightarrow0}  \int_0^{x^1}\frac{\partial(\omega_\epsilon)^\alpha_1}{\partial{x^2}}(s,x^2)ds \ + \  \omega^{\alpha}_2(0,x^2),
\eeq
with $\epsilon$ convergence in $L^1(\mathcal{U})$. Substituting
\beq\nonumber
\frac{\partial(\omega_\epsilon)^\alpha_1}{\partial{x^2}}=\frac{\partial(\omega_\epsilon)^\alpha_2}{\partial{x^1}}+\left(\frac{\partial(\omega_\epsilon)^\alpha_1}{\partial{x^2}}-\frac{\partial(\omega_\epsilon)^\alpha_2}{\partial{x^1}}\right)
\eeq
into (\ref{2dim2}) gives
\beq\nonumber
\left(\frac{\partial y^\alpha}{\partial x^2}- \omega^\alpha_2\right)(x^1,x^2)= \lim_{\epsilon\rightarrow0}\int_0^{x^1}\left(\frac{\partial(\omega_\epsilon)^\alpha_1}{\partial{x^2}}-\frac{\partial(\omega_\epsilon)^\alpha_2}{\partial{x^1}}\right)(s,x^2)ds
\eeq
with convergence pointwise almost everywhere. Now, using that $(\omega_\epsilon)^\alpha$ is parallel in the $x^2$-direction,  
$\frac{\partial(\omega_\epsilon)^\alpha_1}{\partial x^2}=\Gamma^\sigma_{12}(\omega_\epsilon)^\alpha_\sigma$,
we have
\beq\nonumber
\left(\frac{\partial(\omega_\epsilon)^\alpha_1}{\partial{x^2}}-\frac{\partial(\omega_\epsilon)^\alpha_2}{\partial{x^1}}\right)=-\nabla_1(\omega_\epsilon)^\alpha_2,
\eeq
which leads to
\beq \label{2dim3}
\left(\frac{\partial y^\alpha}{\partial x^2}-\omega^\alpha_2\right)(x^1,x^2)=- \lim_{\epsilon\rightarrow0}\int_0^{x^1}\nabla_1(\omega_\epsilon)^\alpha_2(s,x^2)\,ds .
\eeq
The Gr\"onwall estimate (\ref{2dim5_Groenwall}) now implies
\beq \nonumber
\left|\frac{\partial y^\alpha}{\partial x^2}- \omega^\alpha_2\right| (x^1,x^2) \leq K_0 \int_0^{x^1}|x^2|ds  \ \  \ \ \text{a.e.}
\eeq
which implies the sought after Lipschitz estimate (\ref{loc_inertial_Lemma2_eqn1}).

We now prove (\ref{loc_inertial_Lemma2_eqn2}). By the dominated convergence theorem, we conclude that (\ref{2dim3}) implies
\beq \label{2dim3_limit}
\left(\frac{\partial y^\alpha}{\partial x^2}-\omega^\alpha_2\right)(x^1,x^2)=- \int_0^{x^1}\nabla_1\omega^\alpha_2(s,x^2)\,ds .
\eeq
Differentiating (\ref{2dim3_limit}) in $x^1$-direction gives us
\beq\label{2dim4}
\left(\frac{\partial^2 y^\alpha}{\partial x^1\partial x^2} -\frac{\partial \omega^\alpha_2}{\partial x^1}\right)(x^1,x^2)=-\nabla_1\omega^\alpha_2(x^1,x^2),
\eeq
and taking the absolute value, the Gr\"onwall estimate (\ref{2dim5_Groenwall_again}) gives
\beq\nonumber 
\left|\frac{\partial^2 y^\alpha}{\partial x^1\partial x^2}-\frac{\partial \omega^\alpha_2}{\partial x^1}\right|(x^1,x^2)   \leq K_0 \: \big|x^2\big| \ = \ O(|x|),
\eeq
which is the sought after almost everywhere estimate (\ref{loc_inertial_Lemma2_eqn2}) for $j=1$ and $i=2$. 

It remains to prove (\ref{loc_inertial_Lemma2_eqn2}) for $i=j=2$. For this, differentiate (\ref{2dim3}) in the $x^2$ direction to obtain
\beq\nonumber
\left(\frac{\partial^2 y^\alpha}{\partial x^2\partial x^2}-\frac{\partial \omega^\alpha_2}{\partial x^2}\right)(x^1,x^2)=- \frac{\partial}{\partial x^2}  \lim_{\epsilon\rightarrow0} \int_0^{x^1} \nabla_1(\omega_\epsilon)^\alpha_2(s,x^2)\,ds.
\eeq 
Note that taking $\frac{\partial}{\partial x^2}$ as a derivative in the weak sense, we can exchange $\lim_{\epsilon\rightarrow 0}$ and $\frac{\partial}{\partial x^2}$ by the $L^1$ convergence of the integrand.  Thus by (\ref{2dim5}), 
\beq \label{2dim_uniform}
\int_0^{x^1} \frac{\partial}{\partial x^2} \nabla_1(\omega_\epsilon)^\alpha_2(s,x^2)\,ds = \int_0^{x^1} \Big( (\Gamma_\epsilon)^{\sigma}_{2k}\nabla_1(\omega_\epsilon)^\alpha_\sigma + (R_\epsilon)^\sigma_{k12}(\omega_\epsilon)^\alpha_\sigma \Big)(s,x^2)\,ds,
\eeq
which converges uniformly in $x^2$ as $\epsilon \rightarrow 0$, because the right hand side is continuous in $x^2$ and bounded in light of the Gr\"onwall estimate (\ref{2dim5_Groenwall}).  In light of the Groenwall estimate (\ref{2dim5_Groenwall}),  the integrand on the right hand side of (\ref{2dim_uniform}) is in $L^{\infty}$, we conclude that 
\beq \label{2dim7}
\left|\frac{\partial^2 y^\alpha}{\partial x^2\partial x^2}-\frac{\partial \omega^\alpha_2}{\partial x^2}\right|(x^1,x^2)  \leq  K_0\ |x^1|,
\eeq 
which implies the sought after bound (\ref{loc_inertial_Lemma2_eqn2}) for $i=j=2$.
\qed \\

\noindent{\it Proof of Proposition \ref{loc_inertial_Thm_2D}:}
We show that $y^{\alpha}$ defined in (\ref{2dim1}) are locally inertial at $p$.    For this, consider the transformation law for connections
\beq \label{loc_inertial_Thm_techeqn1}
\Gamma^k_{ij} \frac{\partial y^\alpha}{\partial x^k} = \frac{\partial^2y^\alpha}{\partial x^i\partial x^j} +\Gamma^\alpha_{\beta\gamma}\frac{\partial y^\beta}{\partial x^i}\frac{\partial y^\gamma}{\partial x^j}.
\eeq
Combining (\ref{loc_inertial_Lemma1_eqn}) and (\ref{loc_inertial_Lemma2_eqn2}), we obtain
\beq \nonumber
\frac{\partial^2y^\alpha}{\partial x^i\partial x^j} = \Gamma^k_{ij} \omega^\alpha_k + O(|x|).
\eeq
Substituting the previous equation into (\ref{loc_inertial_Thm_techeqn1}) and using that $w^\alpha_k=\frac{\partial y^\alpha}{\partial x^k} + O(|x|)$ by (\ref{loc_inertial_Lemma2_eqn1}), the Christoffel symbols $\Gamma^k_{ij}$ cancel on both sides and we get
\beq \nonumber
 \Gamma^\alpha_{\beta\gamma}\frac{\partial y^\beta}{\partial x^i}\frac{\partial y^\gamma}{\partial x^j} = O(|x|).
\eeq
We then conclude with the sought after estimate (\ref{loc_inertial_Thm_2D}), using that $O(|x|)=O(|y|)$ and that the Jacobians $\frac{\partial y^\beta}{\partial x^i}$ are invertible. 
This completes the proof of Proposition \ref{loc_inertial_Thm_2D}.  \hfill $\Box$
\vspace{.2cm}

Finally, it is interesting to point out what goes wrong in the pursuit of the above construction for locally inertial frames in 3-dimensions. Essentially, the analog of (\ref{loc_inertial_Lemma2_eqn2}) does not hold in 3-dimensions. That is, defining coordinates in analogy to (\ref{2dim1}) leads to 
\beq \label{2dim1_3D}
y^{\alpha}(x^1,x^2,x^3) \equiv \int_0^{x^1}\omega^{\alpha}_1(s,x^2,x^3)ds+\int_0^{x^2}\omega^{\alpha}_2(0,s,x^3)ds + \int_0^{x^3} \omega^\alpha_3(0,0,s)ds,
\eeq
and the analog of (\ref{whatwewant1}) again holds. However, since $\nabla_2 (\omega_\epsilon)_1^\alpha(x^1,x^2,x^3)$ is not zero when $x_3 \neq 0$, we get in (\ref{2dim3}) an additional error function bounded in $L^{\infty}$ which is $O(x^3)$, but whose derivative is not $O(x)$.  That is, we obtain
\beq \label{2dim_problem}
\left(\frac{\partial y^\alpha}{\partial x^2}-\omega^\alpha_2\right)(x^1,x^2)=- \lim_{\epsilon\rightarrow0}\int_0^{x^1} \nabla_1(\omega_\epsilon)^\alpha_2(s,x^2)\; ds \ +\; \int_0^{x^1}  O(x^3) \; ds.
\eeq
Thus, differentiating (\ref{2dim_problem}) in $x^2$ direction in order to mimic the step leading to equation (\ref{2dim_uniform}) above, the derivative falls on the term $\int_0^{x^1}  O(x^3) ds$, the derivative of an $L^\infty$ function, which does not in general produce an error $O(x)$.

\subsection{Proof of Theorem \ref{thm_noRS}}

Assume $\mathcal{M}$ is a spherically symmetric Lorentz manifold, by which we mean that coordinates exist in which the metric tensor takes the  form
\beq \label{SSY_metric}
ds^2 = - A(x^1,x^2) \big(dx^1\big)^2 + 2 E(x^1,x^2)dx^1 dx^2 + B(x^1,x^2) \big(dx^2\big)^2 + C(x^1,x^2) d\Omega^2,
\eeq
where the components $A,B,C$ and $E$ are assumed to be Lipschitz continuous functions, $-(AB+E^2)<0$ and $C>0$.  Here $d\Omega^2 \equiv  d\phi^2 + \sin(\phi)^2 d\theta^2$ is the line element on the unit sphere,  $x^3=\phi\in (0,\pi)$, $x^4=\theta\in (-\pi,\pi)$, and we assume without loss of generality that $(x^1,x^2)$ are centered at $(0,0)$, with $(x^1,x^2)\in(-1,1)\times(-1,1)\equiv\Omega_2$.  (General spherically symmetric Lorentz metrics can generically be transformed to coordinates where the metric takes the form (\ref{SSY_metric}), \cite{Weinberg}.)   Assume further that the metric connection $\Gamma^k_{ij}$ and Riemann curvature tensor $R^k_{lij}$ are bounded in $L^\infty$ in coordinates $(x^1,x^2,\phi,\theta)$.  
 To prove Theorem \ref{thm_noRS}, it suffices to prove that the metric (\ref{SSY_metric}) admits locally inertial coordinates at each point $p\in \Omega_2$ within the atlas of $C^{1,1}$ coordinate transformations, in the sense of Definition \ref{Deflipcont}. 
To this end, we now extend the constructions in Section \ref{susec8.1} to spherically symmetric spacetimes.  We start with the following lemma:

\begin{Lemma} \label{smooth_C_lemma}
Assume the metric (\ref{SSY_metric}) is Lipschitz continuous. If the Einstein tensor of the metric (\ref{SSY_metric}) is bounded in $L^\infty$, then $C(x^1,x^2)\in C^{1,1}(\Omega_2)$.
\end{Lemma}

\proof
An explicit computation of the first three contravariant components of the Einstein tensor $G^{11}$, $G^{12}$ and $G^{22}$,  yields
\begin{eqnarray}\label{EFE in SSY}
\frac{\partial^2 C}{\partial x^2  \partial x^2} &=&\kappa C |g| G^{11}+ l.o.t.,   \cr
\frac{\partial^2 C}{\partial x^1  \partial x^2} &=&- \kappa C |g| G^{12}+ l.o.t.,  \cr
\frac{\partial^2 C}{\partial x^1  \partial x^1} &=&\kappa C |g| G^{22}+l.o.t.,
\end{eqnarray}
where $|g|\equiv  -A B - E^2$ and $l.o.t.$ denotes terms containing only zero and first order metric derivatives, (c.f. MAPLE). From this we can read off the regularity of $C$. Namely, when $G^{\mu\nu} \,\in\, L^\infty$ and the metric is Lipschitz continuous metric, the right hand side of (\ref{EFE in SSY}) is in $L^\infty$. Thus we conclude that second order weak derivatives of $C$ are in $L^{\infty}$, which is equivalent to ~$C\,\in\, C^{1,1}$, (c.f. \cite{Evans}).
\qed \\

In the proof of the theorem to follow, it is interesting to observe that our assumption that the curvature tensor is bounded in $L^{\infty}$ comes in at two different points in the argument to imply the existence of locally inertial frames for (\ref{SSY_metric}) when the connection is only in $L^{\infty}$.   First, we apply Proposition \ref{loc_inertial_Thm_2D} to the $2$-dimensional metric $$ds^2 = - A(x^1,x^2) \big(dx^1\big)^2 + 2 E(x^1,x^2)dx^1 dx^2 + B(x^1,x^2) \big(dx^2\big)^2$$ to obtain coordinates $y^{\alpha}$ in which the connection is Lipschitz continuous at the center point $p=(0,0)$, but only for indices running from $1$ to $2$.  An explicit computation then shows that the remaining components involving the angular indices are in fact Lipschitz continuous, one degree smoother than $L^{\infty}$,  because $C$ is the only differentiated metric component in these connection components, and $C$ is one degree more regular than $A,B$ and $E$, by Lemma \ref{smooth_C_lemma}.   This extra degree of regularity in $C$ is crucial because it ensures that the connection coefficients not addressed by our $2$-dimensional method, must be Lipschitz continuous, as a second consequence of our assumption that the curvature tensor is bounded in $L^{\infty}$.   The resulting Lipschitz continuity of $\Gamma$ at $p$ in $y$-coordinates allows us to introduce a further smooth coordinate transformation, quadratic in $y$,  which breaks the spherical symmetry, and sets the value of the connection to zero at the center point $p$, while preserving the established Lipschitz continuity at $p$ in $y$-coordinates.
\vspace{.2cm}

\noindent{\it Proof of Theorem \ref{loc_inertial_Thm_sph}:}
For $\alpha=1,2$, we introduce the two $1$-forms 
$$
\omega^\alpha = \omega^\alpha_1 dx^1 + \omega^\alpha_2 dx^2 
$$ 
as solutions of $\nabla_2\omega^{\alpha}=0$ with variables $(x^1,x^2)$ assuming $(x^3,x^4)=(\phi_0,\theta_0)$ fixed in $(0,\pi)\times(-\pi,\pi)$. That is, $\omega^1$ and $\omega^2$ are solutions of 
\beq \label{sph_loc_inertial_ODE2}
\frac{\partial\omega^\alpha_j}{\partial x^2}(x^1,x^2) -\Gamma^k_{2j}\omega^\alpha_k(x^1,x^2)=0,
\eeq  
for initial data $\omega^\alpha_j(x^1,0,\phi_0,\theta_0)$ at $x^2=0$ with $\nabla_1 \omega^\alpha_j(x^1,0,\phi_0,\theta_0)=0$ for $j=1,2$, c.f. (\ref{loc_inertial_ODE2}). Since the angular dependence is kept fixed in (\ref{sph_loc_inertial_ODE2}), estimate (\ref{loc_inertial_Lemma1_eqn}) of Lemma \ref{loc_inertial_Lemma1} holds again for both $\omega^\alpha$, that is,
\beq \label{spherical_2dim2a}
\frac{\partial\omega^\alpha_j}{\partial x^i} = \Gamma^k_{ij}\big|_{(\phi_0,\theta_0)} \omega^\alpha_k + O(|x^1|+|x^2|),    
\eeq 
for $ \alpha=1,2$, where $\Gamma^k_{ij}\big|_{(\phi_0,\theta_0)}$ denotes $\Gamma^k_{ij}$ evaluated at fixed $(\phi,\theta)=(\phi_0,\theta_0)$.

Similar to (\ref{2dim1}), we define the function $y^\alpha$ for $\alpha=1,2$ as
\beq \label{spherical_2dim1}
y^{\alpha}(x^1,x^2,\phi,\theta) \equiv \int_0^{x^1}\omega^{\alpha}_1(s,x^2,\phi_0,\theta_0)ds+\int_{0}^{x^2}\omega^{\alpha}_2(0,s,\phi_0,\theta_0)ds,
\eeq
(the right hand side evaluated at fixed values of the angular variables), and set 
$y^3 = \phi$,  $y^4 = \theta.$
Thus estimates (\ref{loc_inertial_Lemma2_eqn1}) and (\ref{loc_inertial_Lemma2_eqn2}) hold,
\begin{eqnarray} \label{spherical_2dim2b}
\frac{\partial y^\alpha}{\partial x^k} &=& w^\alpha_k + O(|x^1|+|x^2|), \cr
\frac{\partial^2y^\alpha}{\partial x^i\partial x^j} &=& \frac{\partial\omega^\alpha_j}{\partial x^i} + O(|x^1|+|x^2|),
\end{eqnarray}
for $i,j,k=1,2$ and $\alpha=1,2$, and where the right hand side is evaluated at the fixed angular values $(\phi_0,\theta_0)$.
Combining estimates (\ref{spherical_2dim2a}) and (\ref{spherical_2dim2b}), we obtain for $\alpha=1,2$
\beq \label{sph_final_estimate}
\frac{\partial^2y^\alpha}{\partial x^i\partial x^j}
=\Gamma^k_{ij}\big|_{(\phi_0,\theta_0)}\frac{\partial y^\alpha}{\partial x^k}+O(|x^1|+|x^2|),\ \  i,j=1,2.
\eeq
To complete the proof, consider again the transformation  
\beq \label{sph_transfo_law}
\frac{\partial y^\alpha}{\partial x^k}\Gamma^k_{ij}=\frac{\partial^2 y^\alpha}{\partial x^i\partial x^j} +\Gamma^\alpha_{\beta\gamma}\frac{\partial y^\beta}{\partial x^i}\frac{\partial y^\gamma}{\partial x^j}.
\eeq
Substituting (\ref{sph_final_estimate}) we obtain
\beq \label{sph_final_estimate_2}
\Gamma^\alpha_{\beta\gamma}\frac{\partial y^\beta}{\partial x^i}\frac{\partial y^\gamma}{\partial x^j} = \Big(\Gamma^k_{ij} - \Gamma^k_{ij}\big|_{(\phi_0,\theta_0)}\Big) \frac{\partial y^\alpha}{\partial x^k} +O(|x^1|+|x^2|) .
\eeq
Using now that the metric and its inverse are smooth in $\phi$ and $\theta$, we can Taylor expand $\Gamma^k_{ij}$ around $(\phi_0,\theta_0)$ to obtain 
$$
\Gamma^k_{ij} - \Gamma^k_{ij}\big|_{(\phi_0,\theta_0)} = O(|\phi-\phi_0| + |\theta-\theta_0|)
$$  
for $i,j=1,2$. Thus, since the Jacobian  $\frac{\partial y^\alpha}{\partial x^k}$ is invertible and since $\frac{\partial y^\alpha}{\partial x^k}=0$ for $\alpha=1,2$ and $k=3,4$, we can write (\ref{sph_final_estimate_2}) as 
\begin{eqnarray}
\Gamma^\alpha_{\beta\gamma} &=& O(|x^1|+|x^2| + |\phi-\phi_0| + |\theta-\theta_0|) \nonumber \\
&=& O(|y^1|+|y^2| + |\phi-\phi_0| + |\theta-\theta_0|), \hspace{.6cm}   \alpha,\beta,\gamma=1,2.\hspace{.6cm} \label{sph_Gamma_estimate_1&2}
\end{eqnarray}
Keeping in mind that $y^1(0,0)=0=y^2(0,0)$ and that $y^3=\phi$ and $y^4=\theta$, this is the desired Lipschitz estimate for $\alpha,\beta,\gamma=1,2$.

We now derive a Lipschitz estimate of the form (\ref{sph_Gamma_estimate_1&2}) for the cases when $\alpha,\beta$ or $\gamma \neq 1,2$. The transformation to the coordinates $y^\alpha$ defined in (\ref{spherical_2dim1}), preserves the spherically symmetric form of the metric representation (\ref{SSY_metric}). We denote the metric in coordinates $y^\alpha$ by
\beq \label{SSY_metric_y}
ds^2 = - A(y^1,y^2) (dy^1)^2 + 2 E(y^1,y^2)dy^1 dy^2 + B(y^1,y^2) (dy^2)^2 + C(y^1,y^2) d\Omega^2 
\eeq
for Lipschitz continuous metric components $A,B,C,E$, generally different from the components in (\ref{SSY_metric}). Computing the Christoffel symbols of (\ref{SSY_metric_y}), we find that the non-zero connection coefficient not subject to the Lipschitz estimate (\ref{sph_Gamma_estimate_1&2}) are given by
\begin{eqnarray} \label{sph_Christoffel_symbols_rest}
\Gamma^1_{33} &=& \frac{B \dot{C} - E C'}{2(AB + E^2)}, \hspace{1cm}  \Gamma^1_{44} = (\sin\phi)^2 \Gamma^1_{33}, \cr
\Gamma^2_{33} &=& \frac{-E \dot{C} - A C'}{2(AB + E^2)}, \hspace{1cm}  \Gamma^2_{44} = (\sin\phi)^2 \Gamma^2_{33}, \cr
\Gamma^3_{13} &=& \frac{\dot{C}}{2C}, \hspace{1cm} \Gamma^3_{23} = \frac{C'}{2C}, \hspace{1cm} \Gamma^3_{44} = -\sin \phi \cos\phi, \cr
\Gamma^4_{14} &=& \frac{\dot{C}}{2C}, \hspace{1cm} \Gamma^4_{24} = \frac{C'}{2C}, \hspace{1cm} \Gamma^3_{44} = \frac{\cos\phi}{\sin \phi},
\end{eqnarray}
where $\dot{C}\equiv \frac{\partial C}{\partial y^1}$ and $C'\equiv \frac{\partial C}{\partial y^2}$. Observe that we only differentiate $C$ in the above coefficient components but we never differentiate $A$, $B$ or $E$. Since $C$ is $C^{1,1}$ regular by Lemma \ref{smooth_C_lemma}, it follows that the components in (\ref{sph_Christoffel_symbols_rest}) are Lipschitz continuous (as long that $\phi\neq 0$). Combining this with the Lipschitz estimate (\ref{sph_Gamma_estimate_1&2}), we conclude that $\Gamma^\alpha_{\beta\gamma}$ is Lipschitz continuous at $p$ in coordinate $y^\alpha$.

The Christoffel symbols in (\ref{sph_Christoffel_symbols_rest}) are generally non-zero since $\dot{C}$ and $C'$ are non-zero.   Since a non-singular coordinate transformation preserving the metric form (\ref{SSY_metric}) cannot map $\dot{C}$ and $C'$ to zero, we need a transformation that breaks the form (\ref{SSY_metric}).   To complete the proof, we now introduce a coordinate transformation which preserves the Lipschitz continuity at $p$ and maps the Christoffel symbols to zero at the point $p$. Without loss of generality, we assume that $y(p)=(0,0,\phi_0,0)$ for some $\phi_0 \in (0,\pi)$. Since the Christoffel symbols in coordinates $y^\alpha$ are Lipschitz continuous at $p$ (and hence defined at $p$), we can introduce (for $\mu=1,...,4$) the smooth coordinate transformation         
\beq \label{sph_final_transfo}
z^\mu(y) \equiv \frac12 \delta^\mu_\alpha \Gamma^\alpha_{\beta\gamma}\big|_p \: y^\beta y^\gamma + \delta^\mu_\alpha \: y^\alpha   + c^\mu_\beta \: y^\beta  + c^\mu
\eeq
where $\delta^\mu_\alpha$ denotes the Kronecker symbol and the constants $c^\mu$ and the constant coefficients $c^\mu_\beta$ are defined by
\begin{eqnarray} \nonumber
c^\mu \equiv -\frac12 \delta^\mu_\alpha \Gamma^\alpha_{33}\big|_p \: \phi_0^{\ 2} - \delta^\mu_3 \phi_0    
\hspace{1cm} \text{and} \hspace{1cm} 
c^\mu_\beta \equiv   -\frac12 \delta^\mu_\alpha \Gamma^\alpha_{\beta 3}\big|_p  \phi_0.
\end{eqnarray}
By our definition of $c^\mu$ and $c^\mu_\beta$, it follows from (\ref{sph_final_transfo}) that 
\beq \label{sph_final_transfo_eqn1}
z(p)=0 \hspace{1cm} \text{and} \hspace{1cm} \frac{\partial z^\mu}{\partial y^\alpha}\Big|_p = \delta^\mu_\alpha. 
\eeq
Moreover, (\ref{sph_final_transfo}) implies that
\beq \label{sph_final_transfo_eqn2}
\frac{\partial^2 z^\mu}{\partial y^\beta \partial y^\gamma}\Big|_p =\delta^\mu_\alpha \Gamma^\alpha_{\beta\gamma}\big|_p.
\eeq
From the transformation law of connections together with (\ref{sph_final_transfo_eqn1}) and (\ref{sph_final_transfo_eqn2}), we find that the Christoffel symbols in coordinates $z^\mu$ vanish at $p$. Namely, (\ref{sph_final_transfo_eqn1}) and (\ref{sph_final_transfo_eqn2}) imply that the transformation law (\ref{sph_transfo_law}) evaluated at $p$ is given by
\begin{eqnarray} \nonumber
\frac{\partial z^\sigma}{\partial y^\alpha} \Gamma^\alpha_{\beta\gamma} 
=\frac{\partial^2 z^\sigma}{\partial y^\beta \partial y^\gamma} +\Gamma^\sigma_{\mu\nu}\frac{\partial z^\mu}{\partial y^\beta}\frac{\partial z^\nu}{\partial y^\gamma} 
=\delta^\sigma_\alpha \Gamma^\alpha_{\beta\gamma} +\Gamma^\sigma_{\mu\nu}\delta^\mu_\beta \delta^\nu_\gamma ,
\end{eqnarray} 
which implies that the Christoffel symbol in coordinates $z^\mu$ satisfies
$
\Gamma^\sigma_{\mu\nu}\big|_p =0,
$
for all $\sigma,\mu,\nu \in \{1,...,4\}$. Clearly, since the transformation is smooth, it preserves the Lipschitz continuity of $\Gamma$ at $p$. Denoting the coordinates $z^j$ by $y^\alpha$, we proved the sought after Lipschitz estimate (\ref{locallyinertial}). This completes the proof of Theorem \ref{loc_inertial_Thm_sph}.
\hfill$\Box$

\section{Conclusion}

We prove that the question of whether there exists a $C^{1,1}$ coordinate transformation which smooths an $L^{\infty}$ symmetric connection $\Gamma$ to $C^{0,1}$ in some neighborhood is equivalent to the existence of a Lipschitz continuous $(1,2)$-tensor $\tilde{\Gamma}$ such that $\Gamma-\tilde{\Gamma}$ is Riemann-flat in that neighborhood.   Somewhat surprisingly, the coordinate construction in the proof of Proposition \ref{thm1} can be modified to give locally inertial frames for Lipschitz metrics, and this applies to solutions of the Einstein-Euler equations generated by the Glimm scheme, \cite{GroahTemple},  but only in spherically symmetric spacetimes.  The $C^{1,1}$ regularity issue regarding whether the metric can be smoothed to $C^{1,1}$, remains open, even in spherical symmetry.  In summary, the space of $L^{\infty}$ connections with $L^{\infty}$ curvature tensors provides a consistent general framework for shock wave theory in General Relativity, and the problem whether {\it weak} regularity singularities exist in spherically symmetric spacetimes, or whether {\it weak} or {\it strong} regularity singularities exist at points of more complicated shock wave interaction, remains an open problem for which the results here provide a new geometric perspective.

\section*{Acknowledgments}

We thank Heinrich Freist\"uhler for a careful reading of our manuscript and very helpful suggestions.

\section*{Funding}

M. Reintjes was funded through CAPES-Brazil as a Post-Doctorate at IMPA (Rio de Janeiro) in 2015 and  2016. In 2017 and 2018, M. Reintjes was supported by FCT/Portugal through (GPSEinstein) PTDC/MAT-ANA/1275/2014  and UID/MAT/04459/2013. B. Temple was supported by NSF Applied Mathematics Grant Number DMS-010-2493.

%

\providecommand{\bysame}{\leavevmode\hbox to3em{\hrulefill}\thinspace}
\providecommand{\MR}{\relax\ifhmode\unskip\space\fi MR }
\providecommand{\MRhref}[2]{%
  \href{http://www.ams.org/mathscinet-getitem?mr=#1}{#2}
}
\providecommand{\href}[2]{#2}

\end{document}